\documentclass[preprint,showpacs,preprintnumbers,amsmath,amssymb]{revtex4}
\usepackage{graphicx}

\begin{document}

\title{Analytic solutions for Baxter's model \\
of sticky hard sphere fluids\\
within closures different from the Percus-Yevick approximation.}
\author{Domenico Gazzillo}
\author{Achille Giacometti}
\affiliation{Istituto Nazionale per la Fisica della Materia and \\
Dipartimento di Chimica Fisica, Universit\`{a} di Venezia, \\
S. Marta DD 2137, I-30123 Venezia, Italy}
\email{gazzillo@unive.it \; achille@unive.it}
\date{\today }

\begin{abstract}
We discuss structural and thermodynamical properties of Baxter's adhesive
hard sphere model within a class of closures which includes the
Percus-Yevick (PY) one. The common feature of all these closures is to have
a direct correlation function vanishing beyond a certain range, each closure
being identified by a different approximation within the original
square-well region. This allows a common analytical solution of the
Ornstein-Zernike integral equation, with the cavity function playing a
privileged role. A careful analytical treatment of the equation of state is
reported. Numerical comparison with Monte Carlo simulations shows that the
PY approximation lies between simpler closures, which may yield less
accurate predictions but are easily extensible to multi-component fluids,
and more sophisticate closures which give more precise predictions but can
hardly be extended to mixtures. In regimes typical for colloidal and protein
solutions, however, it is found that the perturbative closures, even when
limited to first-order, produce satisfactory results.
\end{abstract}

\maketitle

\section{INTRODUCTION}

In spite of their highly idealized character, `sticky hard sphere' (SHS)
models continue to attract considerable attention in the recent literature
on mesoscopic fluids. The reasons are the following. On the one hand, these
models are simple enough to allow analytic solution of the Ornstein-Zernike
(OZ) integral equation of the liquid state theory within some appropriate
approximation. On the other hand, their potentials, albeit crude, exhibit
realistic physical features (repulsion plus surface attraction) typical of
the interactions between particles of many macromolecular fluids.
Furthermore, the SHS models have already been successfully employed in
studies on colloidal suspensions, micelles, protein solutions,
microemulsions, flocculation, and crystallization \cite%
{Stell91,Jamnik96,Santos98,Jamnik01}.

The first and most known model (SHS1) was proposed by Baxter \cite%
{Baxter68,Baxter71a,Baxter71b,Barboy74}. Its pair potential adds to a hard
sphere (HS) repulsion an attractive square-well (SW), which becomes
infinitely deep and narrow according to a limiting procedure (Baxter's
`sticky limit') that keeps the second virial coefficient finite \cite%
{Baxter68}. For pure fluids the starting potential is

\begin{equation}
\beta \phi _{{\rm SW1}}(r)=\left\{ 
\begin{array}{lll}
+\infty , &  & 0<r<\sigma \\ 
-\beta \varepsilon _{{\rm SW1}}, &  & \sigma \leq r\leq R \\ 
0, &  & r>R%
\end{array}%
\right. ,  \label{i1}
\end{equation}

\begin{equation}
\varepsilon _{{\rm SW1}}=k_{B}T\ln \left( \alpha \ \frac{R}{R-\sigma }%
\right) ,\qquad \alpha =\frac{1}{12\tau },  \label{i2}
\end{equation}%
where $r$ is the distance between particles, $\sigma $ the HS diameter, $%
\varepsilon _{{\rm SW1}}$ and $R-\sigma $ denote the depth and width of the
well, respectively, and $\beta =(k_{B}T)^{-1}$ ( $k_{B}$ being Boltzmann's
constant, and $T$ the temperature). The dimensionless parameter $\tau \geq 0$
is an {\it arbitrary} monotonically increasing function of $T~$ \cite%
{Baxter71b}, and $\alpha $ is a measure of `stickiness' between particles.
In the limit $\tau \rightarrow \infty ,$ $\phi _{{\rm SW1}}(r)$ becomes the
HS potential with diameter $R.$ The logarithmic form of $\varepsilon _{{\rm %
SW1}}$ serves to generate a simple expression for the Boltzmann factor, $%
e(r)=\exp \left[ -\beta \phi (r)\right] $, i.e.

\begin{equation}
e_{{\rm SW1}}(r)=\exp \left[ -\beta \phi _{{\rm SW1}}(r)\right] =\left\{ 
\begin{array}{lll}
0, &  & 0<r<\sigma \\ 
\alpha R\ \left( R-\sigma \right) ^{-1}, &  & \sigma \leq r\leq R \\ 
1, &  & r>R%
\end{array}%
\right. ,  \label{i3}
\end{equation}

The `sticky limit' amounts to the limit $R-\sigma \rightarrow 0$, and the
second virial coefficient, $B_{2}=2\pi \int_{0}^{\infty }dr$ $r^{2}\left[
1-e(r)\right] ,$ becomes $B_{2\text{ }}^{{\rm SHS1}}=\lim_{R\rightarrow
\sigma ^{+}}B_{2\text{ }}^{{\rm SW1}}=\left( 2\pi /3\right) \sigma ^{3}\left[
1-\left( 4\tau \right) ^{-1}\right] $.

The OZ equation for the SHS1 model admits analytical solution within the 
{\it Percus-Yevick} (PY) {\it approximation} \cite%
{Baxter68,Perram75,Barboy79}. However, the difficulty in applying the
multi-component SHS1-PY solution to polydisperse colloidal mixtures with a
large number $p$ of components (due to the presence of $p(p+1)/2$ coupled
quadratic equations \cite{Perram75} to be solved numerically) has led to two
more recent attempts of finding an alternative SHS model, which could be
analytically tractable even in the general multi-component case \cite%
{Brey87,Mier89,Ginoza96,Tutschka98,Tutschka00,Tutschka01,Tutschka02,Gazzillo00,Gazzillo02a,Gazzillo02b,Gazzillo03a,Gazzillo03b,Timoneda89,Ginoza01,Jamnik91,Hoye93}%
. We shall refer to them as SHS2 \cite{Brey87,Mier89,Ginoza96} and SHS3
models \cite{Gazzillo03a,Gazzillo03b}. In SHS2 the adhesive part of the
potential is generated from an attractive Yukawa tail, whose amplitude and
inverse range become infinite in the sticky limit, while their ratio remains
constant. The SHS3 model has been proposed to replace SHS2, overcoming a
recently discovered pathology while retaining some useful features of the
previous model.

As remarked in Ref. \cite{Brey87}, the crucial difference between SHS1 and SHS2 
lies in the attractive part of the interaction, since the area of the SHS1 well
vanishes as $(R-\sigma)\ln(R-\sigma)$ when the width $R-\sigma$ goes to zero, while
the corresponding integral remains finite in the SHS2 case. 

For SHS2, the OZ equation was analytically solved within the {\it mean
spherical approximation} (MSA) \cite{Mier89,Ginoza96}, and the
multi-component SHS2-MSA solution is indeed much simpler than its SHS1-PY
counterpart.

On the other hand, Baxter's SHS1 model was never solved analytically within
the MSA or other closures different from the PY one. Note that in the MSA
the main input for the direct correlation function is the potential SW1
itself, which has, in the sticky limit, a weaker (logarithmic) divergence
than the Boltzmann factor. Thus it would be interesting to investigate other
approximations which consider, more correctly, the Boltzmann factor rather
than the potential, in a way similar to the PY case.

Motivated by this scenario, in this paper we would like to revisit Baxter's
model within a new framework. We consider a general class of closures which,
as the PY, hinge on the vanishing of the direct correlation function beyond
a finite range $R$. Each component of the class is identified by a different
approximation to the cavity function in the intermediate region $(\sigma ,R)$%
. We shall show that this difference survives even in the limit $%
R\rightarrow \sigma $ and hence it gives rise to different solutions, each
of \ them being characterized by a different expression for a single
parameter $q_{\sigma }$.

In this class we first consider a subset of closures which is based on
density expansion of the cavity function. All these closures yield a simpler
solution than the PY one, and yet they provide a satisfactory description on
regimes frequently encountered in experiments on colloids and globular
proteins. The solutions corresponding to these closures could be
straightforwardly extended to multi-component fluids, unlike the PY case.

In addition, we identify a second subclass of closures as a possible route
to improve over the PY approximation.

Finally, we provide a careful discussion on the equation of state derived
from this class of solutions. In this process, we critically discuss the
correct way of taking the `sticky limit' and its influence on the predicted
structural and termodynamic properties of the model. A short-cut, but not
fully equivalent, procedure may in fact generate an incorrect result for the
virial equation of state (EOS). We point out that pitfalls of this sort have
already occurred in the case of SHS1 mixtures \cite{Barboy79,Tutschka01}.
Such a problem should be clearly borne in mind when investigating
thermodynamic properties of more complex fluid models which include sticky
interaction terms in the pair potential.

\section{BASIC\ THEORY}

\subsection{Integral equations}

For a one-component fluid of $N$ molecules in a volume $V,$ with spherically
symmetric interactions, the OZ equation is the convolution relationship

\begin{equation}
h(r)=c(r)+\rho \left( c\ast h\right) (r)\equiv c(r)+\gamma \left( r\right) ,
\label{ie1}
\end{equation}%
which defines the `direct correlation function' (DCF) $c(r)$ in terms of the
`total correlation function' $h(r)$, deviation of the `radial distribution
function' (RDF) $g(r)$ from its ideal gas value of $1$, i.e. $h(r)=g(r)-1$.
Here, the asterisk represents three-dimensional convolution, i.e. $\left(
c\ast h\right) (r)\equiv \int d{\bf r}^{\prime }$\ $c(r^{\prime })h\left(
\left| {\bf r-r}^{\prime }\right| \right) $. Moreover, $r$ refers to the
distance between particles, $\rho =N/V$ is the number density, and the
convolution term $\gamma \left( r\right) $ is the `indirect correlation
function', continuous with its first three derivatives everywhere, even when
the potential has a hard-core part. The OZ equation can be cast into the
Baxter form \cite{Baxter71a}%
\begin{eqnarray}
rc\left( r\right) &=&-q^{\prime }(r)+2\pi \rho \int_{0}^{\infty }du\ q\left(
u\right) q^{\prime }\left( r+u\right) ,  \label{ie2a} \\
rh\left( r\right) &=&-q^{\prime }(r)+2\pi \rho \int_{0}^{\infty }du\ q\left(
u\right) \left( r-u\right) h\left( |r-u|\right) ,  \label{ie2b}
\end{eqnarray}%
where the `factor correlation function' $q(r)$ is an auxiliary function, and
the prime denotes differentiation with respect to $r$. Solving the Baxter
equations is tantamount to determining $q(r)$, from which $c\left( r\right) $
and $h\left( r\right) $ can be easily calculated. To solve the OZ equation,
some approximate `closure' relating $c\left( r\right) $, $h\left( r\right) $
and the potential $\phi \left( r\right) $ must be added.

\subsection{Closures}

For hard-core fluids,\ whose $\phi \left( r\right) $ may be written as the
sum of a HS contribution and a tail, i.e., $\phi \left( r\right) =\phi _{%
{\rm HS}}\left( r\right) +\phi _{{\rm tail}}\left( r\right) $ (with $\phi _{%
{\rm HS}}\left( r\right) =+\infty $ for $r<\sigma ,$ and $=0$ for $r\geq
\sigma ),$ an approximation is required only for $r\geq \sigma .$ In fact,
in this case both $c(r)$ and $g(r)$ are exactly defined inside the core,
i.e.,

\begin{eqnarray}
g(r) &=&0,\qquad 0<r<\sigma ,  \label{clos0d} \\
c(r) &=&-\left[ 1+\gamma \left( r\right) \right] ,\qquad 0<r<\sigma .
\label{clos0e}
\end{eqnarray}

Let us consider some known approximations to the exact DCF:

i) the {\it mean spherical approximation} (MSA),

\begin{equation}
c_{{\rm MSA}}\left( r\right) =-\beta \phi \left( r\right) ,\qquad r\geq
\sigma .  \label{clos1}
\end{equation}

ii) A {\it modified mean spherical approximation} (mMSA) \cite%
{Gazzillo03b,Huang84,Pini02}

\begin{equation}
c_{{\rm mMSA}}\left( r\right) =f\left( r\right) ,\qquad r\geq \sigma .
\label{clos2}
\end{equation}%
where $f\left( r\right) =e(r)-1$ is the Mayer function.

iii) The {\it Percus-Yevick} (PY) closure,

\begin{equation}
c_{{\rm PY}}\left( r\right) =f(r)\left[ 1+\gamma \left( r\right) \right]
,\qquad r\geq 0.  \label{clos3}
\end{equation}

It is well known that the MSA assumption is rather accurate at large $r$
values, but not in the region just outside the hard core. For short-ranged
potentials, one thus expects that the MSA performance could be improved by
using the mMSA which replaces $-\beta \phi \left( r\right) $ with $f\left(
r\right) $. Moreover, the mMSA yields the correct zero-order term in the
density expansion of the DCF, since $c_{{\rm exact}}\left( r\right)
\rightarrow f\left( r\right) $ when $\rho \rightarrow 0$.

Note that, when applied to Baxter's SW1 potential, all these closures imply
that $c(r)$ vanishes beyond the range of the potential. In the present paper
we will thus focus, for this model, on the class of {\it all} closures which
can be written as%
\begin{equation}
c_{{\rm GPY}}(r)=\left\{ 
\begin{array}{cc}
-\left[ 1+\gamma \left( r\right) \right] , & \qquad 0<r<\sigma , \\ 
c_{{\rm shrink}}\left( r\right) , & \qquad \sigma <r<R, \\ 
0, & \qquad \text{\ }r>R.%
\end{array}%
\right.   \label{clos0f}
\end{equation}%
We shall refer to this expression as {\it generalized PY }(GPY) {\it %
approximation.} We remark that in Eq. ($\ref{clos0f}$) the novelty lies
on the fact that it does not
univocally define a single closure, since different approximations can be
obtained by different definitions of $c_{{\rm shrink}}\left( r\right) $, the
DCF in the shrinking interval ($\sigma ,R$). As it turns out, 
these differences persist {\it %
even in the sticky limit} $R\rightarrow \sigma $!

Strictly speaking, all closures of the GPY class, when different from the PY
one, are {\it mixed closures}, being a combination of the PY approximation
for $r>R$ with a different approximation to $c(r)$ for $\sigma <r<R$. For
instance, $c_{{\rm shrink}}\left( r\right) =-\beta \phi _{{\rm tail}}\left(
r\right) $ yields a MSA/PY closure, while one
recovers the PY approximation (PY/PY) when $c_{{\rm shrink}}\left( r\right)
=f(r)\left[ 1+\gamma \left( r\right) \right] $. We recall that combinations
of closures, such as hypernetted chain - mean spherical (HNC/MSA) and
Percus-Yevick - mean spherical (PY/MSA) approximations have already been
employed in studies on electrolytes with sticky interactions \cite%
{Rasaiah85,Herrera91}.

\subsection{General analytic solution}

Let us apply the GPY approximation to the model defined by Baxter's SW1
potential. Different approximations to $c_{{\rm shrink}}\left( r\right) $
generate, of course, different solutions (which we are not able to write
explicitly in analytic terms). It is however surprising that, as shown in
Appendix A, the differences among all these solutions
drastically reduce in the sticky limit 
all being included in one common parameter, $%
q_{\sigma }$, while the $r$ dependence becomes identical. As a consequence
of this degeneracy, the OZ equation for the SHS1 model admits, for 
{\it all} members of the GPY class of (mixed) closures, the following
general {\it analytic} solution 
\begin{equation}
q_{{\rm SHS1-GPY}}(r)=\left[ \frac{1}{2}a(r^{2}-\sigma ^{2})+b\sigma
(r-\sigma )+q_{\sigma }\sigma ^{2}\right] \theta \left( \sigma -r\right)
,\qquad \text{ for \ }r\geq 0,  \label{so1}
\end{equation}%
\begin{equation}
a=\frac{1+2\eta }{\left( 1-\eta \right) ^{2}}-\frac{12q_{\sigma }\eta }{%
1-\eta },\qquad b=-\frac{3\eta }{2\left( 1-\eta \right) ^{2}}+\frac{%
6q_{\sigma }\eta }{1-\eta },  \label{so2}
\end{equation}%
\begin{equation}
q_{\sigma }=\frac{1}{12}\lambda =\frac{1}{12\tau }\ y_{{\rm SHS1-GPY}%
}(\sigma )\equiv \frac{1}{12\tau }\lim_{R\rightarrow \sigma }y_{{\rm SW1-GPY}%
}(\sigma ^{+})  \label{so3}
\end{equation}%
where $y_{{\rm SHS1-GPY}}(\sigma )$ denotes the cavity function at contact
and, for simplicity, the subscript SHS1-GPY has been omitted in $a$, $b$,
and $q_{\sigma }$.

The GPY solution includes and extends the SHS1-PY expression found by Baxter %
\cite{Baxter71a}, and formally resembles the MSA one for the SHS2 model \cite%
{Ginoza96}. In fact, as regards the dependence on $r$, the functional form
is identical in both the mentioned cases. The
distinguishing feature of each element of the GPY class lies, however, 
in the parameter dependence of 
$q_{\sigma }(\eta ,\tau )$, that mantains a
`memory' of the situation before the sticky limit is taken,
depending upon the closure chosen for $c_{{\rm shrink}}\left( r\right) $
(and upon the model) \cite{note0}.

Note that the sticky limits of $y_{{\rm SW1-GPY}}(\sigma ^{+})$ and $y_{{\rm %
SW1-GPY}}(R^{+})$ are in general different, since different approximations
to $y(r)$ may be used for $\sigma <r<R$ and $r>R$, respectively.

The function $q_{{\rm SHS1-GPY}}(r)$ is discontinuous at $r=\sigma ,$ since $%
q_{{\rm SHS1-GPY}}(\sigma ^{-})\neq q_{{\rm SHS1-GPY}}(\sigma ^{+})=0,$ and
this jump leads to $q_{{\rm SHS1}}^{\prime }(r)=\left[ ar+b\sigma \right]
\theta \left( \sigma -r\right) -q_{\sigma }\sigma ^{2}\delta (r-\sigma )$.
Different approximations correspond to different expressions for $q_{\sigma
}=q_{{\rm SHS1-GPY}}(\sigma ^{-})/\sigma ^{2}$.

It is also of some interest to write down an explicit expression for the DCF
inside the core, which can be derived by using Eq. ($\ref{ie2a}$), and\ reads%
\begin{eqnarray}
c_{{\rm SHS1-GPY}}(r) &=&\left\{ -12\eta \ q_{\sigma }^{2}\ \frac{\sigma }{r}%
-a^{2}+12\eta \left[ \frac{1}{2}(a+b)^{2}-aq_{\sigma }\right] \frac{r}{%
\sigma }-\frac{\eta }{2}a^{2}\left( \frac{r}{\sigma }\right) ^{3}\right\}
\theta (\sigma -r)  \nonumber \\
&&+\ q_{\sigma }\ \sigma \delta (r-\sigma ),\qquad \qquad 0\leq r<\sigma .
\label{so4}
\end{eqnarray}%
Each solution $q_{{\rm SHS1}}(r)$ of the GPY class is fully characterized by
its approximation to $y_{{\rm SHS1}}(\sigma )$. In the following we thus
discuss some particular solutions focusing on the cavity function.

\subsection{Cavity function}

The cavity function $y(r)$ is defined by%
\begin{equation}
g(r)=e(r)y(r).  \label{y1}
\end{equation}%
Sometimes it is also expressed as

\begin{equation}
y(r)=1+\gamma \left( r\right) +d(r),  \label{y4}
\end{equation}%
which defines the `tail function' $d(r)$ \cite{Katsov00,Stell63}, involved
also in the exact relationship

\begin{equation}
c(r)=f(r)\left[ 1+\gamma \left( r\right) \right] +e(r)d(r).  \label{y3}
\end{equation}%
Note that the PY approximation amounts to set $d(r)=0$ $\ \forall r\geq 0.$

The exact $y(r)$ is a well-defined continuous function of $r$ with at least
2 continuous derivatives, even when the potential $\phi \left( r\right) $
itself has a hard-core region or other discontinuities \cite{Katsov00}.
Since in Baxter's model the cavity function at the contact $y(\sigma )$
plays a crucial role in characterizing a particular approximation within the
GPY class, it is important to know, for SHS1, the first three coefficients
of the expansion 
\begin{equation}
y_{{\rm exact}}(\sigma )=y_{0}+y_{1}\eta +y_{2}\eta ^{2}+\cdots  \label{cexp}
\end{equation}%
( $\eta =\rho v_{0}$, $v_{0}=(\pi /6)\sigma ^{3}$ being the packing fraction
and the particle volume, respectively), which can be derived from the SHS1 
{\it exact} virial coefficients \cite{Post86} (see Section V.B and Eq. ($\ref%
{ae0b}$)). One finds

\begin{equation}
\left\{ 
\begin{array}{c}
y_{0}=1, \\ 
\text{\ }y_{1}=\allowbreak \frac{5}{2}-t+\frac{1}{12}\allowbreak t^{2}, \\ 
\text{\ }y_{2}=4.\,\allowbreak 591\,-4.\,\allowbreak 076\ t+1.\,\allowbreak
518\ \allowbreak t^{2}-0.232\ t^{3} \\ 
\qquad \quad +1.\,\allowbreak 056\times 10^{-2}\allowbreak \
t^{4}+1.\,\allowbreak 302\,5\times 10^{-4}\ t^{5},%
\end{array}%
\right.  \label{y5}
\end{equation}%
where%
\begin{equation}
t\equiv \tau ^{-1}.
\end{equation}

Note that, within the GPY class, {\it mixed }closures imply that $y_{{\rm GPY%
}}(r)=1+\gamma \left( r\right) =y_{{\rm PY}}(r)$ when $r>\sigma $, whereas $%
\ y_{{\rm GPY}}(r)$ for $\sigma <r<R$ may largely differ from the PY
expression.

A second crucial remark is that, while $y_{{\rm exact}}(r)$ is continuous
everywhere, a cavity function derived from an approximate closure may be
discontinuous. This drawback occurs, for instance, in the MSA (see Eq. ($\ref%
{ymsa}$) below) when the potential has some discontinuities outside the
core. In the case of SW fluid, $y_{{\rm MSA}}(r)$ may thus be considered
continuous at $r=\sigma $, but it is discontinuous at $r=R$ (a fact which
has remarkable consequences on the derivation of the corresponding MSA
virial equation of state \cite{Smith77}).

\subsection{Correct way of taking the sticky limit}

Before ending this Section, we would like to stress also a general
methodological point related to the way the sticky limit is performed, in
view of its importance in the evaluation of the virial EOS. The correct
procedure - hereafter referred to as {\it sticky limit at the end} {\bf - }%
involves three steps:{\bf \ }first, solve the OZ equation, within some
approximate closure, for the `starting' potential $\phi _{{\rm SW1}}(r)$.
Second, compute structural and thermodynamic properties of the SW1 fluid.
Finally, perform the sticky limit.

In Baxter's case, an immediate difficulty arises from the lack of analytic
solution for SW fluids. Remarkably, one can find the correct SHS1-PY
solution even in the absence of a complete solution of the SW1-PY model, by
using perturbative expansions in powers of $R-\sigma $. The starting point is

\begin{equation}
g_{{\rm SW1}}(r)=\left[ \theta \left( r-R\right) +\ \alpha \ \frac{R}{%
R-\sigma }\theta \left( r-\sigma \right) \ \theta \left( R-r\right) \right]
\ y_{{\rm SW1}}(r),  \label{sl1}
\end{equation}%
where the expression ($\ref{i3}$) for $e_{{\rm SW1}}(r)$ has been inserted
into Eq. ($\ref{y1}$)$.$ Note that no approximation is contained in Eq. ($%
\ref{sl1}$), if $y_{{\rm SW1}}(r)$ is exact. The use of $y_{{\rm PY}%
}(r)=1+\gamma (r)$ yields $g_{{\rm SW1-PY}}(r)$. On inserting this
expression into Eq. ($\ref{ie2b}$), one can determine the necessary
properties of \ the unknown $q_{{\rm SW1-PY}}(r)$ (see Appendix A) and
hence, in the sticky limit, of $q_{{\rm SHS1-PY}}(r)$.

However, the evaluation of structural and thermodynamic properties involves
operations such as derivatives and integrals, which do not necessarly
commute with the sticky limit. An underestimation of this fact has led to
the existence of some incorrect results in the literature when the sticky
limit was carried out already at the level of the Boltzmann factor ({\it %
sticky limit from the outset}), i.e.

\begin{equation}
e_{{\rm SHS1}}(r)\equiv \lim_{R\rightarrow \sigma ^{+}}e_{{\rm SW1}%
}(r)=\theta \left( r-\sigma \right) +\ \alpha \ \sigma \delta _{+}\left(
r-\sigma \right) ,  \label{sl2}
\end{equation}%
which implies that \cite{note1} 
\begin{equation}
g_{{\rm SHS1}}(r)=\theta \left( r-\sigma \right) \ y_{{\rm SHS1}}(r)+\
\alpha \ \sigma y_{{\rm SHS1}}(\sigma )\ \delta _{+}\left( r-\sigma \right) ,
\label{sl3}
\end{equation}

\begin{equation}
c_{{\rm SHS1}}(r)=\alpha \ \sigma y_{{\rm SHS1}}(\sigma )\ \delta _{+}\left(
r-\sigma \right) ,\text{ \ \ \ for \ \ }r\geq \sigma  \label{sl3b}
\end{equation}%
(note that in Eq. ($\ref{sl2}$) one cannot write $e_{{\rm SHS1}}(r)=\exp %
\left[ -\beta \phi _{{\rm SHS1}}(r)\right] $, since $\phi _{{\rm SHS1}%
}(r)\equiv \lim_{R\rightarrow \sigma }\phi _{{\rm SW1}}(r)\ $diverges
logarithmically). This short-cut procedure of `sticky limit from the outset'
is rather common in the literature on systems with sticky terms in the pair
potential (not only SHS fluids). We stress that this approach, albeit
computationally simpler, is not always equivalent to the `sticky limit at
the end' (which remains the unique correct procedure) and may be dangerous
in calculations of thermodynamic properties.

In view of this scenario, we have performed an accurate re-analysis of
Baxter's work and found that the way of taking the sticky limit does not
affect $q(r)$, as well as any quantity computed from it without $r$%
-derivatives, i.e., for instance, the structure factor $S(k),$ the
compressibility and energy EOS's, the internal energy and the Helmholtz free
energy. This is {\it not} the case of the virial EOS as it will be shown
later.

Being aware of the danger hidden in the procedure of `sticky limit from the
outset', Baxter used it only whenever possible \cite{Baxter71a}. In the
present paper, we follow only the surely correct procedure of `sticky limit
at the end'.

\section{PARTICULAR SOLUTIONS}

In addition to the PY approximation, we now consider other GPY solutions
relevant to {\it mixed }closures $\cdots $/PY, where the dots refer to the
approximation chosen for $c_{{\rm shrink}}(r)$, the DCF in the shrinking
interval ($\sigma $,$R$).

\subsection{MSA/PY or MSA closure}

Let us start from $c_{{\rm shrink}}(r)=-\beta \phi _{{\rm tail}}\left(
r\right) $. For any square-well potential, and in particular for Baxter's
SW1, this MSA/PY mixed closure coincides with the pure MSA one. In terms of
cavity function, one can write

\begin{equation}
y_{{\rm MSA}}\left( r\right) =\exp \left[ \beta \phi _{{\rm tail}}\left(
r\right) \right] \ \left[ 1+\gamma \left( r\right) -\beta \phi _{{\rm tail}%
}\left( r\right) \right] \text{ \qquad for \ }r>\sigma ,  \label{ymsa}
\end{equation}%
and thus

\[
y_{{\rm SW1-MSA}}\left( r\right) =\left\{ 
\begin{array}{cc}
\frac{R-\sigma }{\alpha R}\left[ 1+\gamma _{{\rm SW1-MSA}}\left( r\right)
+\ln \frac{\alpha R}{R-\sigma }\right] , & \qquad \sigma <r<R, \\ 
1+\gamma _{{\rm SW1-MSA}}\left( r\right) , & r>R,%
\end{array}%
\right. 
\]%
which is clearly discontinuous at $r=R$. In the $R\rightarrow \sigma $ limit
one gets:

\begin{equation}
y_{{\rm SHS1-MSA}}(\sigma )=0,  \label{b3}
\end{equation}%
since the factor $1+\gamma _{{\rm SHS1-MSA}}\left( \sigma \right) $ remains
finite, while the remaining term  in $y_{{\rm SW1-MSA}}\left( \sigma \right) 
$, involving the product of the well-width $R-\sigma $ times the potential
at contact, vanishes due to the weaker (logarithmic) divergence of  $%
-\beta \phi _{{\rm SW1}}\left( \sigma \right) $ (on the contrary, $R-\sigma $
times the Boltzmann factor would yield a non-zero finite result). As a
consequence, one finds

\begin{equation}
q_{\sigma }^{{\rm MSA}}=0,\quad \Longrightarrow \quad q_{{\rm SHS1-MSA}%
}(r)=q_{{\rm HS-PY}}(r).  \label{b4}
\end{equation}%
This means that {\it the SHS1-MSA factor correlation function differs
strongly from its SHS2-MSA counterpart and coincides with the HS-PY one. }
In other words, for Baxter's potential, the MSA is a far too poor
approximation, unable to include into $q\left( r\right) $ any effect due to
the surface adhesion.

On the other hand, we can anticipate that the SHS1-MSA virial pressure does
not coincide with its HS-PY counterpart, but diverges! 
As discussed in detail later on (Section V.C and Appendix B),
this is due to the discontinuity at $r=R$ in the cavity function $y_{{\rm SW1-MSA}}(r)$
within the MSA approximation. Hence a particularly careful treatment of the jump in the
cavity function, appearing in the virial equation of state, is required 
in this case \cite{Smith77}, unlike all those closures which preserve the continuity
of $y(r)$ (e.g. PY).
As a result, we find the somewhat paradoxical situation of having the  limit of the 
factor correlation function $q_{{\rm SHS1-MSA}}(r)$ identical with the
hard sphere one $q_{{\rm HS-PY}}(r)$, whereas the corresponding limits of the virial 
equations of state are different.
This unexpected result may appear puzzling at first sight. However we attribuite it 
to the strong thermodynamic inconsistency generated by the MSA in the sticky limit.

\subsection{mMSA closure or C0-approximation}

Henceforth the mMSA will be denoted as C0{\it -approximation}, to emphasize
that it corresponds to the {\it exact} \ zero-order approximation, $c_{{\rm %
shrink}}(r)=f\left( r\right) $, for the DCF in the shrinking interval
outside the core.

Now $y_{{\rm C0}}\left( r\right) =\ 1+\gamma \left( r\right) \exp \left[
\beta \phi _{{\rm tail}}\left( r\right) \right] $ \ for $r>\sigma $, and

\[
y_{{\rm SW1-C0}}\left( r\right) =1+\frac{R-\sigma }{\alpha R}\ \gamma _{{\rm %
SW1-C0}}\left( r\right) ,\qquad \sigma <r<R. 
\]%
This implies that

\begin{equation}
y_{{\rm SHS1-C0}}(\sigma )=1,  \label{yc0}
\end{equation}

\begin{equation}
q_{\sigma }^{{\rm C0}}=\frac{t\ }{12}\ ,  \label{b6}
\end{equation}%
which depends on temperature, but is {\it density-independent.}

It is interesting to remark that the resulting $q_{{\rm SHS1-C0}}(r)${\it \ }%
solution coincides with the mMSA solution of the SHS3 model \cite%
{Gazzillo03b}, provided that the reduced temperature $T^{\ast \text{ }}$ of
that potential is replaced with Baxter's $\tau $. It is also worth noting
that the SHS3-mMSA solution is formally coincident with the SHS2-MSA one %
\cite{Gazzillo03a,Gazzillo03b}. As a consequence of this mapping, most of
the MSA results obtained in the past for SHS2 (which turns out to be {\it %
ill-defined} from a thermodynamical point of view) are found to still hold
within a correct framework \cite{Gazzillo03b}.

\subsection{C1-approximation}

The most natural extension of the previous approximation is the inclusion of
the next order correction in the density expansion. We shall denote this as
C1{\it -approximation},{\it \ } \ which is defined by 
\begin{equation}
c_{{\rm shrink}}(r)=f\left( r\right) \left[ 1+\gamma _{1}\left( r\right)
\rho \right] ,
\end{equation}%
where $\gamma _{1}\left( r\right) =\ \left( f\ast f\right) (r)$ is the
first-order coefficient in the density expansion of $\gamma \left( r,\rho
\right) $. This approximation is {\it exact up to first-order} in density.
As a consequence, we find

\begin{equation}
y_{{\rm SHS1-C1}}(\sigma )=1+y_{1}(t)\ \eta ,
\end{equation}%
with $y_{1}$ being the exact first-order coefficient given by Eq. ($\ref{y5}$%
). The corresponding $q_{\sigma }^{{\rm C1}}$ follows immediately.

Clearly, higher orders terms in the density expansion of $c_{{\rm shrink}%
}(r) $ could be considered. However, even the inclusion of the next order
correction would require a much more elaborate analysis compared to the
first-order one. Furthermore it is reasonable to expect that higher order, $%
t $-dependent, terms should be introduced in a way balanced with the HS
part, in view of the slow convergence of the perturbative series. For these
reasons, in this paper, our analysis will be limited to the first correction
only.

\subsection{Baxter's PY solution}

Next we consider the PY closure, which cannot be deduced from C0 by adding a
finite number of perturbative terms.

In this case the approximation to $y(r)$ is `continuous', in the sense that
the same closure is employed both in $(\sigma ,R)$ and $(R,+\infty )$. Eq. ($%
\ref{so3}$) then\ becomes

\begin{equation}
q_{\sigma }^{{\rm PY}}=\frac{t}{12}\ \ y_{{\rm SHS1-PY}}(\sigma ),
\label{b1}
\end{equation}%
where the contact value of the PY cavity function can be determined
analytically by solving a quadratic equation, usually written in terms of $%
\lambda $. Only the smaller of the two real solutions (when they exist) is
physically significant \cite{Baxter68}. The result reads%
\begin{equation}
y_{{\rm SHS1-PY}}(\sigma )=\frac{y_{0}(\sigma )}{\frac{1}{2}\left[ 1+\frac{%
\eta }{1-\eta }t+\sqrt{\left( 1+\frac{\eta }{1-\eta }t\right) ^{2}-\frac{1}{3%
}\eta \ y_{0}(\sigma )\ t^{2}}\ \right] },  \label{b2}
\end{equation}%
where $y_{0}(\sigma )$ is the HS-PY result 
\begin{equation}
y_{{\rm HS-PY}_{{\rm V}}}(\sigma )=\left( 1+\eta /2\right) /(1-\eta )^{2}.
\label{b5}
\end{equation}%
It is interesting to compare the first terms of the expansion of $y_{{\rm %
SHS1-PY}}(\sigma )$ in powers of $\eta $ with the exact results given by Eq.
($\ref{y5}$). One finds that $y_{0}^{{\rm SHS1-PY}}=1=y_{0}^{{\rm exact}}$, $%
y_{1}^{{\rm SHS1-PY}}=y_{1}^{{\rm exact}}$, while

\begin{eqnarray}
y_{2}^{{\rm SHS1-PY}} &=&\allowbreak 4-\frac{7}{2}t+\frac{17}{12}t^{2}-\frac{%
1}{4}t^{3}+\frac{1}{72}t^{4}  \nonumber \\
&=&\allowbreak 4.0-3.\,\allowbreak 5\ t+1.\,\allowbreak 417\ \allowbreak
t^{2}-0.25\ t^{3}+1.\,\allowbreak 389\times 10^{-2}\allowbreak \ t^{4}.
\end{eqnarray}%
Note that in $y_{2}^{{\rm exact}}$ there is a $t^{5}$-term, missing here.$%
\allowbreak $

\subsection{Beyond the PY closure}

By exploiting the Baxter-OZ equations, one can evaluate $1+\gamma _{{\rm %
SHS1-GPY}}(\sigma )$ and find (see Appendix B)%
\begin{equation}
\gamma _{{\rm SHS1-GPY}}(\sigma )=y_{{\rm HS-PY}_{{\rm V}}}(\sigma )-1-\frac{%
\eta }{1-\eta }\lambda +\frac{1}{12}\eta \lambda ^{2}.  \label{e1}
\end{equation}%
An improvement over the PY approximation can be obtained by recalling Eq. ($%
\ref{y4}$) and writing%
\begin{eqnarray}
y_{{\rm SHS1-GPY}}(\sigma ) &=&1+\gamma _{{\rm SHS1-GPY}}(\sigma )+d(\sigma )
\nonumber \\
&=&y_{{\rm HS-PY}_{{\rm V}}}(\sigma )-\frac{\eta }{1-\eta }\ t\ y_{{\rm %
SHS1-GPY}}(\sigma )+\frac{1}{12}\eta \left[ t\ y_{{\rm SHS1-GPY}}(\sigma )%
\right] ^{2}+d(\sigma ).  \label{so6}
\end{eqnarray}%
We stress that in Eq.(\ref{so6}) the presence of $\ d(\sigma )$ is our key
for improving over the PY closure. If we are able to give an explicit
approximation for $d(\sigma )$, then $y_{{\rm SHS1-GPY}}(\sigma )$ can be
computed (analytically or numerically) from Eq. ($\ref{so6}$).

In the following we shall attempt to motivate a particular form of $d(\sigma
)$ using a combination of exact heuristic arguments. Our guidance in this
task hinges on two observations. First, we note that density expansion of
Eq. (\ref{y4}) suggests that $d(\sigma )$ starts with a $\rho ^{2}$-term,
and hence it is negligible at low density. The PY case indeed corresponds to
set $d(\sigma )=0$.

On the other hand, it is well known that the PY approximation yields a
description of the cavity function for HS which is far from being fully
accurate. This can be quickly established from the density expansion of Eq. (%
\ref{b5}) which yields $y_{2}^{{\rm HS-PY_{V}}}=4$, markedly different from $%
y_{2}^{{\rm HS-exact}}=4.591$ (see Eq. (\ref{y5}) ). In order to cope with
both the above remarks, we propose the following simple expression for $%
d(\sigma )$, which will be referred to as GPY1,

\begin{equation}
d_{{\rm GPY1}}(\sigma )=\frac{\eta ^{2}}{2\left( 1-\eta \right) ^{3}}+\left(
-\frac{t}{2}+\frac{t^{2}}{12}\right) \ \allowbreak \eta ^{2},  \label{gpy1}
\end{equation}%
Substitution in Eq. (\ref{e1}) yields Eq. (\ref{b2}) again, but now with 
\begin{eqnarray}
y_{{\rm 0}}(\sigma ) &=&\ y_{{\rm CS}}(\sigma )-(t/2-t^{2}/12)\ \eta ^{2}, \\
y_{{\rm CS}}(\sigma ) &=&\left( 1-\eta /2\right) \ /\ (1-\eta )^{3},
\label{y0}
\end{eqnarray}%
The HS part of $d_{{\rm GPY1}}(\sigma )$ has been chosen in order to get $y_{%
{\rm CS}}(\sigma ),$ the well-known Carnahan-Starling expression, which is
more accurate than $y_{{\rm HS-PY}_{{\rm V}}}(\sigma )$\cite{Henderson75}.
On the other hand, the $t$-dependent part of $d_{{\rm GPY1}}(\sigma )$
ensures that now the $\eta -$expansion of $y_{{\rm SHS1-GPY1}}(\sigma )$
yields%
\begin{equation}
y_{2}^{{\rm SHS1-GPY1}}=4.5-4\ t+1.5\ t^{2}-\frac{1}{4}t^{3}+\frac{1}{72}%
t^{4},
\end{equation}%
where the first three terms are closer to the exact ones than their PY
counterparts.

In spite of the fact that the approximation given in Eq. (\ref{gpy1}) is not
fully physically motivated, it nevertheless does represent an improvement
over the PY one, as it will be shown in Section VI where numerical tests are
reported.

\section{STRUCTURE}

The structure of the fluid is described by $g(r)$, or by the structure
factor, 
\begin{equation}
S(k)=1+\rho \widetilde{h}\left( k\right) =\left[ 1-\rho \widetilde{c}\left(
k\right) \right] ^{-1}=\left[ \widehat{Q}\left( -k\right) \widehat{Q}\left(
k\right) \right] ^{-1},  \label{sfact}
\end{equation}%
where $k$ is the magnitude of the scattering wave vector,$\ \widetilde{h}%
\left( k\right) $ and $\widetilde{c}\left( k\right) $ denote the
three-dimensional Fourier transform of $h\left( r\right) $ and $c(r)$,
respectively, while $\widehat{Q}(k)=1-2\pi \rho \ \widehat{q}\left( k\right)
,$ with $\widehat{q}\left( k\right) $ being the unidimensional Fourier
transform of $q\left( r\right) $. Using Eq. (\ref{so1}), the calculation of $%
S(k)$ is straightforward and yields

\begin{equation}
S_{{\rm SHS1-GPY}}(k\sigma )=\frac{1}{\left[ 1-12\eta \ \widehat{q}_{{\rm %
\cos }}\left( k\sigma \right) \right] ^{2}+\left[ 12\eta \ \widehat{q}_{\sin
}\left( k\sigma \right) \right] ^{2}},
\end{equation}%
where 
\begin{eqnarray}
\widehat{q}_{{\rm \cos }}\left( x\right) &=&\frac{1}{2}a\ I_{2}\left(
x\right) +b\ I_{1}\left( x\right) +q_{0}\ I_{0}\left( x\right) ,  \nonumber
\\
\widehat{q}_{\sin }\left( x\right) &=&\frac{1}{2}a\ J_{2}\left( x\right) +b\
J_{1}\left( x\right) +q_{0}\ J_{0}\left( x\right) ,
\end{eqnarray}%
\begin{equation}
q_{0}=q(0)/\sigma ^{2}=-\ \left[ 2\left( 1-\eta \right) \right]
^{-1}+q_{\sigma }\,,
\end{equation}%
and \cite{Regnaut89} 
\begin{equation}
I_{n}\left( x\right) =\int_{0}^{1}du\ u^{n}\cos (xu),\qquad J_{n}\left(
x\right) =\int_{0}^{1}du\ u^{n}\sin (xu).
\end{equation}%
\qquad

\section{THERMODYNAMICS}

The thermodynamic properties can be obtained through the usual
compressibility, energy and virial routes to the EOS. In any approximate
theory, however, these results do not agree with each other, due to
thermodynamic inconsistency.

The {\it exact} values of the first virial coefficients in the $\eta $%
-expansion of the SHS1 pressure 
\begin{equation}
\beta Pv_{0}=\eta +B_{2}\eta ^{2}+B_{3}\eta ^{3}+B_{4}\eta ^{4}+\cdots ,
\end{equation}
are \cite{Post86}

\begin{equation}
\left\{ 
\begin{array}{c}
B_{2}=4-t, \\ 
B_{3}=10-5t+t^{2}-\frac{1}{18}t^{3}, \\ 
B_{4}=18.36477-13.77358\ t+6.114\ t^{2}-1.518\ t^{3} \\ 
\qquad \qquad \qquad +0.17398\ t^{4}-6.33514\times 10^{-3}\
t^{5}-6.51271\times 10^{-5}\ t^{6}%
\end{array}%
\right.
\end{equation}

In the following we will discuss results for the MSA, C0 and C1 closures,
and compare them with the PY ones. Thermodynamics of the GPY1 approximation
will not be considered in this paper.

\subsection{Compressibility route}

In general, the {\it compressibility }$(C)$ {\it equation }reads

\begin{equation}
\left( \frac{\partial \beta P}{\partial \rho }\right) _{T}=1-\rho \widetilde{%
c}\left( 0\right) =\widehat{Q}^{2}\left( 0\right) =a^{2}=S^{-1}(0)
\label{eosc}
\end{equation}%
where the factorization given in Eq.(\ref{sfact}) as well as the definition
of $a$ have been exploited. Integration with respect to density then yields
the compressibility EOS, i.e. for $Z\equiv \beta P/\rho $

\begin{equation}
Z_{C}\ =\eta ^{-1}\int_{0}^{\eta }a^{2}\ d\eta .
\end{equation}%
For our model and within the GPY class of closures, Eqs. ($\ref{so2}$) yields

\begin{equation}
a_{{\rm SHS1-GPY}}^{2}=\frac{\left( 1+2\eta \right) ^{2}}{\left( 1-\eta
\right) ^{4}}-\frac{2\eta \left( 1+2\eta \right) }{\left( 1-\eta \right) ^{3}%
}\ \lambda (\eta ,t)+\left( \frac{\eta }{1-\eta }\right) ^{2}\ \lambda
^{2}(\eta ,t).
\end{equation}

In particular, for the MSA one has the simple result

\begin{equation}
Z_{C}^{{\rm SHS1-MSA}}\ =Z_{C}^{{\rm HS-PY}}=\ \frac{1+\eta +\eta ^{2}}{%
(1-\eta )^{3}}.
\end{equation}%
which is consistent with that reported in Eq. (\ref{b4}).

Then, for the $c_{0}${\it -approximation }(mMSA) compressibility EOS one
finds%
\begin{eqnarray}
Z_{C}^{{\rm SHS1-C0}}\ \eta &=&\beta P_{C}^{{\rm SHS1-C0}}\ v_{0}=t^{2}\
\eta +\left( 4t+2t^{2}\right) \ln (1-\eta )  \nonumber \\
&&+\left( 1+4t+t^{2}\right) \frac{\eta }{1-\eta }  \nonumber \\
&&+3\left( 1-t\right) \left( \frac{\eta }{1-\eta }\right) ^{2}+3\left( \frac{%
\eta }{1-\eta }\right) ^{3}.  \label{zc0}
\end{eqnarray}%
while the $c_{1}${\it -approximation }gives a more complicated expression

\begin{eqnarray}
Z_{C}^{{\rm SHS1-C}\text{1}}\ \eta &=&Z_{C}^{{\rm SHS1-C0}}\ \eta  \nonumber
\\
&&+\left( 4+4t+3\lambda _{1}\right) \ \lambda _{1}\eta +\left( t+\lambda
_{1}\right) \ \lambda _{1}\eta ^{2}+\frac{1}{3}\allowbreak \lambda
_{1}{}^{2}\eta ^{3}  \nonumber \\
&&+\left( 14+6t+4\lambda _{1}\right) \ \lambda _{1}\ln \left( 1-\eta \right)
\nonumber \\
&&+\left( 10+2t+\lambda _{1}\right) \ \lambda _{1}\frac{\eta }{1-\eta }%
-3\lambda _{1}\left( \frac{\eta }{1-\eta }\right) ^{2},  \label{p1}
\end{eqnarray}%
where $\lambda _{1}\equiv t\ y_{1}(t)$. These results can be compared with
the well known PY ones%
\begin{equation}
Z_{C}^{{\rm SHS1-PY}}=Z_{C}^{{\rm HS-PY}}-\eta \ \frac{(1+\eta /2)}{(1-\eta
)^{2}}\ \lambda +\frac{1}{36}\eta ^{2}\ \lambda ^{3}\ .  \label{zc4}
\end{equation}%
Explicit numerical tests will be reported in Section VI.

It is worthwhile noticing that use of Eq. ($\ref{yc0}$) into Eq. ($\ref{zc4} 
$) does {\it not }yield back Eq. ($\ref{zc0}$).

\subsection{Energy route}

The Helmholtz free energy $A$ can be computed from the cavity function at
contact. In fact, following Baxter \cite{Baxter71b}, it can be shown that

\begin{equation}
\beta \left( A_{{\rm SHS1}}-A_{{\rm HS}}\right) /N=-\eta \ \int_{0}^{t}\ y_{%
{\rm SHS1}}(\sigma )\ dt.  \label{ae0}
\end{equation}%
We stress that this relationship is {\it exact}, independent of any closure.
Once that $A$ is known, the energy EOS can be determined via $Z_{E}=\eta \
\partial (\beta A/N)/\partial \eta $.

Inserting the analogue of the $\eta $-expansion Eq. (\ref{cexp}) for $y_{%
{\rm SHS1}}(\sigma )$ into Eq. ($\ref{ae0}$), one gets the exact
relationship between virial coefficients and $y_{n}(t)$, namely 
\begin{equation}
B_{n}-B_{n}^{{\rm HS}}=-\left( n-1\right) \int_{0}^{t}du\ y_{n-2}(u),\qquad
n\geq 2,  \label{ae0b}
\end{equation}%
which we have exploited to calculate the exact first $y_{n}(t)$'s from the
exact first $B_{n}$'s given by Post and Glandt \cite{Post86}.

>From the contact values calculated in Section III.B one finds

\begin{equation}
\beta A_{{\rm SHS1-MSA}}/N=\beta A_{{\rm HS}}/N
\end{equation}

\begin{equation}
\beta A_{{\rm SHS1-C0}}/N=\beta A_{{\rm HS}}/N-t\ \eta ,  \label{ae1}
\end{equation}

\begin{equation}
\beta A_{{\rm SHS1-C}\text{1}}/N=\beta A_{{\rm HS}}/N-t\ \eta +\left(
-5t+t^{2}-\frac{1}{18}t^{3}\right) \frac{\eta ^{2}}{2},  \label{ae2}
\end{equation}%
and, the corresponding expression for $Z_E$ are: 
\begin{equation}
Z_{E}^{{\rm SHS3-C0}}=Z_{{\rm HS}}-t\ \eta .  \label{ze1}
\end{equation}

\begin{equation}
Z_{E}^{{\rm SHS3-C}\text{1}}=Z_{{\rm HS}}-t\ \eta +\left( -5t+t^{2}-\frac{1}{%
18}t^{3}\right) \eta ^{2},  \label{ze2}
\end{equation}%
Within the PY approximation, analytic expressions for $A_{{\rm SHS3-PY}}$
and $Z_{E}^{{\rm SHS3-PY}}$ are also available \cite{Barboy79}.

It is worth noting that the C0-energy results are mean-field ones: $Z_{E}^{%
{\rm SHS3-C0}}$ predicts only the second virial coefficient correctly,
whereas its higher-order virial coefficients coincide with the HS ones. On
the other hand, the C1 and PY closures give also the exact $B_{3}$.

\subsection{Virial route}

The determination of the virial{\it \ }$(V)$ EOS

\begin{equation}
Z_{V}=1+\frac{2\pi }{3}\rho \int_{0}^{\infty }dr\ r^{3}g(r)\left[ -\beta
\phi ^{\prime }(r)\right] .  \label{eosv}
\end{equation}%
is the most delicate one. The PY result is \cite{Baxter68,Baxter71a}

\begin{eqnarray}
Z_{V}^{{\rm SHS1-PY}} &=&\frac{1}{(1-\eta )^{2}}\left[ 1+2\eta +3\eta ^{2}-%
\frac{2}{3}(1+5\eta )\lambda _{0}+\frac{1}{3}\lambda _{0}^{2}\right] 
\nonumber \\
&&-\frac{\eta }{3\tau (1-\eta )^{3}}\ \left[ 1-5\eta -5\eta ^{2}+6\eta
\lambda _{0}-\lambda _{0}^{2}+\frac{1}{24}\eta ^{-1}\lambda _{0}^{3}\right]
\ .  \label{pyv}
\end{eqnarray}%
with the shorthand $\lambda _{0}=\lambda _{{\rm PY}}\eta (1-\eta ).$

As remarked in Section II.E, a special care must be exerted in evaluating
the sticky limit of this case. As no explicit derivation of Eq. (\ref{pyv})
has been reported in the literature, to the best of our knowledge, we shall
discuss this point in some detail here.

The starting point is the exact expression \cite{Yuste93a,Yuste93b},

\begin{equation}
Z_{V}^{{\rm SHS1}}=\lim_{R\rightarrow \sigma }Z_{V}^{{\rm SW1}}=1+4\eta
\left\{ \ y_{{\rm SHS1}}(\sigma )-\ \alpha \left[ 3y_{{\rm SHS1}}(\sigma
)+\sigma \ \lim_{R\rightarrow \sigma }y_{{\rm SW1}}^{\prime }(\sigma )\right]
\right\} .  \label{swvir}
\end{equation}%
The evaluation of the last term is the {\it crucial} point. To our
knowledge, Bravo Yuste and Santos \cite{Santos98,Yuste93a,Yuste93b} were the
first to clearly stress that, in general, $\lim_{R\rightarrow \sigma }y_{%
{\rm SW1}}^{\prime }(\sigma )$ cannot be replaced with $y_{{\rm SHS1}%
}^{\prime }(\sigma )=[\lim_{R\rightarrow \sigma }y_{{\rm SW1}}^{\prime
}(r)]_{r=\sigma }$ , the analogous quantity corresponding to the procedure
of `sticky limit from the outset', since sticky limit and $r-$%
differentiation do not commute (see Appendix B). In fact, on replacing the
last term in Eq. (\ref{swvir}) with $y_{{\rm SHS1}}^{\prime }(\sigma ),$ one
obtains an incorrect virial EOS, which differs from Eq. ($\ref{pyv}$) (in
particular, the $\lambda _{0}^{3}$-term is missing).

In some sense, a `memory' of the original potential is retained even after
the sticky limit has been performed. The impossibility of getting the
correct derivative from $y(r)$ once the sticky limit has been taken was also
recognized by Seaton and Glandt \cite{Seaton87}.

In short, {\it the correct PY virial EOS cannot be obtained if the sticky
limit is taken `from the outset'. }On the other hand, no such drawbacks
occur when evaluating other structural or thermodynamic properties in the
absence of $r-$differentiation. This important point has sometimes been
overlooked in the past literature. For instance, we mention here the
extension of Eq. ($\ref{pyv}$) to mixtures as proposed by Barboy and Tenne %
\cite{Barboy79}. Their multi-component PY virial equation of state, based
upon the `sticky limit from the outset', is not correct, since it does not
reduce to Eq. (\ref{pyv}) in the one-component limit, as it should. A
similar pitfall appears in the virial pressure of Ref. \cite{Tutschka01}.

In Appendix B it is also shown that

\begin{equation}
Z_{V}^{{\rm SHS1-MSA}}=\lim_{R\rightarrow \sigma }Z_{V}^{{\rm SW1-MSA}%
}=-\infty ,
\end{equation}%
i.e., in the sticky limit the MSA virial EOS differs from $Z_{V}^{{\rm HS-PY}%
}$ and diverges.

Although this result may appear odd at first sight, it is nevertheless
consistent with our previous discussion. The virial EOS depends not only on $%
q(r)$ after the sticky limit, but also on the `tail' of the original
potential.

It is also worth stressing that the evaluation of the MSA virial EOS for SW
fluids embodies a further subtle point not present in the PY virial EOS,
because of the discontinuity of $y_{{\rm SW1-MSA}}(r)$ \cite{Smith77} (see
Appendix B).

One could reasonably expect the above divergence to be a consequence of the
poorness of the MSA closure, which should disappear by resorting to more
accurate approximations. Unfortunately, the virial EOS diverges within the
C0-approximation too (see Appendix B):

\begin{equation}
Z_{V}^{{\rm SHS1-C0}}=\lim_{R\rightarrow \sigma }Z_{V}^{{\rm SW1-C0}%
}=-\infty ,
\end{equation}%
and thus higher-order closures should be considered.

\section{NUMERICAL\ RESULTS}

In order to give a flavor of the difference appearing in the different
approximations, we present here some results and an interesting comparison
with Monte Carlo (MC) simulation data \cite{Kranendonk88,Miller03}.

\subsection{Closures simpler than PY}

Figures 1 and 2 depict MSA, C0, C1 and PY results for the structure factor $%
S(k)$, compared with the corresponding MC data \cite{Kranendonk88,Miller03}.
Four thermodynamic states are considered and we have displayed the
corresponding results in order of decreasing $\tau $. The simulation data
for $S(k)$ stem from a recent improved re-analysis \cite{Miller03} of the
calculations by Kranendonk and Frenkel \cite{Kranendonk88}. These data are
less susceptible to finite size effects than the original ones \cite%
{Kranendonk88}.

As MSA curves coincide with the HS ones, one can clearly appreciate the
effects of surface attraction present in the SHS1 model. As is known \cite%
{Gazzillo00,Gazzillo02a}, at high values of \ $\tau $ repulsive forces
predominate and the structure factor of sticky hard spheres resembles that
of hard spheres, although heights and positions of peaks are quite different
(see Figure 1). Lowering $\tau $ (i.e. decreasing temperature or increasing
stickiness) leads to a different behaviour characterized by a strong
increase of $S(k=0)$, which is related to the isothermal compressibility and
the density fluctuations. The observed development of such a new peak at $%
k=0 $ (see Figure 2) signals the approach to a gas-liquid phase transition ($%
S(0) $ diverges at the spinodal curve). While there is a good overall
agreement between MC data and PY predictions, one can observe that the C0
and C1 curves lie rather close to the PY ones. In addition, in Figures 1a),
1b) and 2c) one clearly sees an improvement in going from the C0 to the C1
approximation, whereas in Fig. 2d) the C0 results are nearly coincident with
the PY ones, at least in the considered interval of parameters. In Figure 2
the largest discrepancies among the different approximations occur near the
origin ( see 2c) ).

Figures 3 and 4 illustrate the behaviour of the compressibility pressure, $%
\beta P_{C}\sigma ^{3}$. In all the considered cases, the performance of the
C0 and C1 approximations is quite good (again, C1 is slightly better than
C0). The C0, C1 and PY results are nearly coincident in the interval $0\leq
\eta \lesssim 0.15$, which covers the concentration range typical of many
colloidal or protein solutions.

Figures 5 and 6 show the predictions for pressure obtained from the energy
route, $\beta P_{E}\sigma ^{3}$. Now, there are much larger differences with
\ respect to the PY and MC results. It is somewhat surprising that, unlike
the PY case \cite{Frenkel03}, within the C0 and C1 closures the
compressibility route appears to be superior than the energy one.

\subsection{Closures beyond PY}

As a final illustrative example, we report here a sample of GPY1 results. In
Figure 7 GPY1 and PY predictions for $S(k)$ are compared with the MC
simulation for the state $(\tau ,\eta )=(0.2,0.32)$. In general the PY
theory tends to underestimate the amplitude of the oscillations in $S(k)$.
Now the improvement present in $y_{2}^{{\rm GPY1}}$ is reflected in the
behaviour of $S(k)$ near the origin. In this region, while the PY curve lies
below the MC data, the GPY1 one has a more correct trend. On the other hand,
since the PY critical point evaluated via the compressibility route ( i.e.
via \ $S(0)$ ) is known to be underestimated with respect to the exact one %
\cite{Frenkel03}, one can reasonably expect that $S_{{\rm SHS1-PY}}(0)$ may
be underestimated too.

We have not considered thermodynamics corresponding to the GPY1 closure,
since such a task would probably require analytical/numerical calculations
which go beyond the scope of the present paper. One could expect, however,
results not significantly different from the PY ones.

\section{CONCLUSIONS}

We have shown that Baxter's model (hard spheres with an infinitely deep and
narrow attractive square-well) is analytically solvable not only within the
PY approximation but also within a general class of closures. We have
denoted this class as `generalized PY' approximation, since all its elements
are `mixed closures', which coincide with the PY one beyond $R$, the upper
boundary of the well, but may employ a different approximation within the
well region $(\sigma ,R)$.

It has been argued that, even in the limit of vanishing width of the well,
there is a `memory' of the approximation to $y(r)$ chosen within the well,
and this has been identified as the distinguishing feature of each closure
of the GPY class.

In particular, we have shown that: i) the SHS1 model can be solved
analytically within the MSA closure, but its SHS1-MSA solution has nothing
to do with the SHS2-MSA one and is equal to $q_{{\rm HS-PY}}(r)$; ii) an
interesting analytic solution can be obtained also within the mMSA (or
C0-approximation). The resulting factor correlation function $q_{{\rm SHS1-C0%
}}(r)$ is formally identical with $q_{{\rm SHS3-mMSA}}(r)$ (if $\tau $ is
replaced with the reduced temperature $T^{\ast }$), obtained for the SHS3
model \cite{Gazzillo03a,Gazzillo03b}.

Although the solution $q(r)$ of the OZ integral equation, and all properties
which can immediately derived from it, can be carried out within the unified
GPY treatment, the calculation of other properties may require an analysis
starting from the original SW potential, i.e. {\it before} the sticky limit.
A detailed analysis of the non-commutativity of the sticky limit with other
operations in some instances, notably for the virial EOS, has been provided.
In principle, one could extend the above remarks from the SHS case to any
general model with sticky terms in the pair potential. Within this more
general viewpoint, the presence of Dirac $\delta $-functions in any of the
functions $f(r)$, $c(r)$, $g(r)$ could be regarded as an indication that the
procedure of `sticky limit from the outset' is implicitly adopted. The
experience of the present paper on SHS systems suggests that special care
should be exerted in evaluating properties where $r$-differentiation is
required, as in the virial EOS, since the corresponding results might be
altered by the non-commutativity of the sticky limit.

The GPY class of closures may be roughly split into two subclasses having
the PY approximation as a separating borderline. The first subclass is
originated by the MSA, which is commonly believed as one of the simplest
possible closures in view of its mean-field character. The main drawback of
this closure lies in its linear relationship with the potential, which
prevents a correct zero-density limit of the correlation functions. We have
cured this by modifying the closure so that the C0-direct correlation
function is related to the Mayer function rather than to the potential, and
in this case the correct zero-density limit is recovered. Finally, an
explicit density-dependence can be perturbatively included in the closure,
order by order, through a density expansion. We have discussed in detail the
analytical results stemming from the zero (C0) and first (C1) order terms in
density expansion and a comparison with Monte Carlo simulations shows that,
in spite of the crudity of the approximations, the resulting structure
factors are rather close to the MC data for a wide range of the parameters.

Our main interest in the closures of this first subclass is that they are
simple enough to allow an easy extension to mixtures and application to
polydisperse fluids, unlike the PY case. The main disadvantage is, on the
other hand, represented by strong thermodynamic inconsistencies, which
cannot be coped with in a simple way.

The closures of the second subclass have in fact been devised to the aim of
reducing these inconsistencies, by exploiting the additional contribution $%
d(\sigma )$ to $y(\sigma )$. We have given only an illustrative example
(called GPY1) to show how small differences in the contact value of the
cavity function can produce, for instance, a different behaviour of $S(k)$
near the origin. On the other hand, the {\it exact} DCF is expected to
possess also a `tail' for $r>R$, which all GPY closures neglect. The
presence of this tail (which could be approximated, for instance, by a
Yukawa function) implies that the {\it exact }solution $q_{{\rm exact}}(r)$
cannot coincide with $q_{{\rm GPY}}(r)$. These slight differences can be
expected to be essential in order to achieve thermodynamic consistency, so
that the use of the mentioned $d(\sigma )$ is useful but probably not
sufficient. In this framework, it would be interesting to compare our $%
d(\sigma )$-based closures with the completely different approach by Santos 
{\it et al.} \cite{Santos98,Yuste93b}, who proposed, for SHS1, to go beyond
the PY closure through a rational function approximation.

Finally, we wish to mention that work along the lines laid down in the
present paper is in progress to investigate, by the multi-component
extension of the C0-approximation, stability boundaries and percolation
threshold of polydisperse colloidal fluids.

\begin{acknowledgments}
 We are particularly grateful to Daan Frenkel and Mark
Miller for having provided us with their very accurate Monte Carlo results
before publication \cite{Miller03}. The Italian MIUR (Ministero
dell'Istruzione, dell'Universit\`{a} e della Ricerca), the INFM (Istituto
Nazionale di Fisica della Materia) are gratefully acknowledged for financial
support.
\end{acknowledgments}

\appendix

\section{Method of solution}

The condition $c(r)=0$ when $r>R$ implies that $q(r)=0$ for $r>R$, and $%
q(R)=0$. Thus, Eq. ($\ref{ie2b}$) may be cast in the form 
\begin{equation}
G\left( r\right) +q^{\prime }(r)=ar+b\sigma +2\pi \rho \left[ \int_{0}^{\min
(R,r-\sigma )}du\ q\left( u\right) G\left( r-u\right) -\int_{\sigma
+r}^{R}du\ q\left( u\right) G\left( u-r\right) \right] ,  \label{ie5}
\end{equation}%
where the definitions $G(r)=rg(r)$, $a=1-2\pi \rho \int_{0}^{\infty }du\
q(u) $, $b\sigma =2\pi \rho \int_{0}^{\infty }du\ uq(u)$, as well as the
core condition, Eq. ($\ref{clos0e}$), have been employed. Here, $\min
(R,r-\sigma )$ denotes the smaller between $R$ and $r-\sigma $.

>From Eqs. ($\ref{ie5}$), one finds that $q(r)$ splits into three parts: $%
q_{1}(r)$ for $0<r<R-\sigma ,$ $q_{2}(r)$ for $R-\sigma <r<\sigma ,$ and $%
q_{3}(r)$ for $\sigma <r<R.$ The function $q(r)$ is continuous at $%
r=R-\sigma ,$ $\sigma ,$ and $R,$ while $q_{2}(r)$ is always represented by
the second degree polynomial $\frac{1}{2}a(r^{2}-\sigma ^{2})+b\sigma
(r-\sigma )+q(\sigma )$ \cite{Barboy79}.

As $R-\sigma \rightarrow 0$, the intervals ($0,R-\sigma $) and ($\sigma ,R$)
vanish, and $q(r)$ reduces to the sticky limit of $q_{2}(r).$ To obtain $a_{%
{\rm SHS1}}\equiv \lim_{R\rightarrow \sigma }a,$ as well as $b_{{\rm SHS1}},$
it is indifferent to take the sticky limit `at the end' or `from the outset'
(see Section II.E), since such a limit commutes with the $u$-integration
included in the definition of $a$ and $b$. Because of the continuity
condition $q_{2}(\sigma ^{-})=q_{3}(\sigma ^{+}),$ the sticky limit of $%
q(\sigma )$ can be evaluated from $q_{3}(r).$ Although the explicit solution
is not known, integrating Eq. ($\ref{ie5}$) yields a formally exact
expression of $q_{3}(r),$ from which we get

\[
q_{3}(\sigma ^{+};R)=\frac{1}{2}a(\sigma ^{2}-R^{2})+b\sigma (\sigma -R)+
F_{1}(\sigma ;R)-2\pi \rho \ F_{2}(\sigma ;R) , 
\]

\[
F_{1}(\sigma ;R)=\int_{\sigma }^{R}dt\ G(t)=G(\sigma ^{+})\left( R-\sigma
\right) +\frac{1}{2}G^{\prime }(\sigma ^{+})\left( R-\sigma \right)
^{2}+\cdots , 
\]%
\[
F_{2}(\sigma ;R)=\int_{\sigma }^{R}ds\ \int_{\sigma }^{s}dt\ q_{1}(s-t)G(t)=%
\frac{1}{2}q_{1}(0^{+})G(\sigma ^{+})\left( R-\sigma \right) ^{2}+\cdots .\ 
\]
It is reasonable to assume that $a$ and $b$ tend to finite values, and the
same occurs for $q_{1}(0^{+})$, whereas $G(\sigma ^{+})$ and $G^{\prime
}(\sigma ^{+})$ may diverge. Consequently, $q_{{\rm SHS1}}(\sigma )$ is
determined by

\[
\lim_{R\rightarrow \sigma ^{+}}q_{3}(\sigma ^{+};R)=\lim_{R\rightarrow
\sigma ^{+}}\int_{\sigma }^{R}dt\ G(t)=\lim_{R\rightarrow \sigma ^{+}}G_{%
{\rm shrink}}(\sigma ^{+};R)(R-\sigma ), 
\]%
since all remaining terms vanish as $R\rightarrow \sigma $ ($G_{{\rm shrink}%
}(\sigma ^{+};R)$ could also be replaced with $C_{{\rm shrink}}(\sigma
^{+};R),$ since the difference becomes zero in the sticky limit). Using Eq. (%
$\ref{y1}$), along with Eq. ($\ref{i3}$), then yields $G_{{\rm SW1}}(\sigma
^{+};R)=\ \alpha \ \sigma \ y_{{\rm SW1}}(\sigma ^{+};R)\ \frac{R}{R-\sigma }
$, and Eq. ($\ref{so3}$) follows immediately. That simple general
relationship allows to determine $q(r)$ of the SHS1 model for {\it any}
closure with $c(r)=0$ when $r>R$.

\section{The virial EOS}

1) {\it Exact case and PY approximation.}

{\it \ }For any SW potential, the virial theorem is awkward since both $g(r)$
and $\phi (r)$ are discontinuous at $r=\sigma $ and $r=R$. Usually, one
rewrites $g(r)\left[ -\beta d\phi (r)/dr\right] $ as $y(r)\ de(r)/dr$ in Eq.
($\ref{eosv}$), and uses

\[
\frac{de_{{\rm SW}}(r)}{dr}=e^{\beta \varepsilon }\ \delta (r-\sigma
)+(1-e^{\beta \varepsilon })\delta (r-R), 
\]

\noindent with $-\varepsilon $ being the well depth. If $y(r)$ is continuous
at both $r=\sigma $ and $r=R$ (as normally occurs), then one obtains

\begin{eqnarray}
Z_{V}^{{\rm SW}} &=&1+\frac{2\pi }{3}\rho \left[ e^{\beta \varepsilon
}\sigma ^{3}\ y_{{\rm SW}}(\sigma )+(1-e^{\beta \varepsilon })R^{3}y_{{\rm SW%
}}(R)\right]  \label{a2} \\
&=&1+4\eta \left\{ g_{{\rm SW}}(\sigma ^{+})+\left( R/\sigma \right) ^{3}\ %
\left[ g_{{\rm SW}}\left( R^{+}\right) -g_{{\rm SW}}(R^{-})\right] \right\} ,
\nonumber
\end{eqnarray}

\noindent which holds for any SW fluid and contains no approximations. For
the particular SW of Baxter's model, the previous exact EOS becomes

\[
Z_{V}^{{\rm SW1}}=1+\frac{2\pi }{3}\rho \left\{ R^{3}y_{{\rm SW1}}(R)-\alpha 
\frac{R}{R-\sigma }\left[ R^{3}y_{{\rm SW1}}(R)-\sigma ^{3}y_{{\rm SW1}%
}(\sigma )\right] \right\} . 
\]%
\ As the exact (as well as the PY) $y(r)$ and $y^{\prime }(r)$ are finite
and continuous everywhere, one can expand $r^{3}y_{{\rm SW1}}(r)$ about $%
r=\sigma $ 
\[
r^{3}y_{{\rm SW1}}(r)=\sigma ^{3}y_{{\rm SW1}}(\sigma )+[r^{3}y_{{\rm SW1}%
}(r)]_{r=\sigma }^{\prime }\ (r-\sigma )+{\cal O}((r-\sigma )^{2}). 
\]

\noindent Evaluating this expression at $r=R,$ inserting it into $Z_{V}^{%
{\rm SW1}},$ and finally taking the sticky limit, one gets \cite%
{Yuste93a,Yuste93b}

\begin{eqnarray}
Z_{V}^{{\rm SHS1}} &=&\lim_{R\rightarrow \sigma }Z_{V}^{{\rm SW1}}=1+\frac{%
2\pi }{3}\rho \left\{ \sigma ^{3}y_{{\rm SHS1}}(\sigma )-\alpha \ \sigma \
\lim_{R\rightarrow \sigma }[r^{3}y_{{\rm SW1}}(r)]_{r=\sigma }^{\prime
}\right\}  \nonumber \\
&=&1+4\eta \left\{ \left( 1-3\alpha \right) y_{{\rm SHS1}}(\sigma )-\alpha \
\sigma \ \lim_{R\rightarrow \sigma }[y_{{\rm SW1}}(r)]_{r=\sigma }^{\prime
}\right\} .
\end{eqnarray}%
Once more, this result is {\it exact, independent of any closure.}

In contrast to what happens for one-dimensional systems of sticky hard rods %
\cite{Yuste93a}, the order of sticky limit and differentiation is essential
in the three-dimensional case, since these operators do not commute now,
i.e.,

\[
\lim_{R\rightarrow \sigma }\left( \lim_{r\rightarrow \sigma }\frac{d}{dr}%
\right) y_{{\rm SW1}}(r)\neq \left( \lim_{r\rightarrow \sigma }\frac{d}{dr}%
\right) \lim_{R\rightarrow \sigma }y_{{\rm SW1}}(r). 
\]

\noindent This is true at least in the PY approximation, where one finds
that \cite{Yuste93a}:

\begin{equation}
\lim_{R\rightarrow \sigma }[y_{{\rm SW1-PY}}(r)]_{r=\sigma }^{\prime }=y_{%
{\rm SHS1-PY}}^{\prime }(\sigma )-\allowbreak \frac{1}{24}\eta ^{2}\lambda
^{3}.  \label{a4}
\end{equation}

To highlight the source of such a difference, we give a detailed
demonstration of this result. Let us define $Y(r)\equiv ry(r),$ and consider
its PY approximation, $Y_{{\rm PY}}(r)=r\left[ 1+\gamma (r)\right] .$ Since $%
\gamma \left( r\right) $ equals $h\left( r\right) -c\left( r\right) $, this
function can be obtained by subtracting Eq. ($\ref{ie2a}$) from Eq. ($\ref%
{ie2b}$). After expanding all integrals $\int_{\sigma }^{R}dt\ldots $ in
powers of $R-\sigma $ and using the relationship $q^{\prime }(\sigma
^{+})=\left( a+b\right) \sigma -G(\sigma ^{+})$, one takes the sticky limit
and obtains%
\begin{equation}
Y_{{\rm SHS1-PY}}(\sigma )=\left( a+b\right) \sigma +12\eta \ \alpha
Y(\sigma )\ q(0)\ /\sigma ^{2},
\end{equation}

\begin{equation}
\lim_{R\rightarrow \sigma }Y_{{\rm SW1-PY}}^{\prime }(\sigma )=a-12\eta \ %
\left[ \left( a+b\right) \ q(0)-\alpha Y(\sigma )\ \lim_{R\rightarrow \sigma
}q_{{\rm SW1-PY}}^{\prime }(0)\ \right] /\sigma ^{2}
\end{equation}%
(for simplicity, the subscript SHS1-PY has been omitted in both right hand
sides).

To evaluate $\lim_{R\rightarrow \sigma }q_{{\rm SW1-PY}}^{\prime }(0),$ one
gets $q_{{\rm SW1-PY}}^{\prime }(0)$ from Eq. ($\ref{ie5}$) for $%
0<r<R-\sigma $. After expanding $\int_{\sigma }^{R}dt\ q\left( t\right)
G\left( t\right) $ up to the $\left( R-\sigma \right) ^{2}$-term (again with
the help of $q^{\prime }(\sigma ^{+})=\left( a+b\right) \sigma -G(\sigma
^{+})$ ), the sticky limit can be performed and yields

\[
\lim_{R\rightarrow \sigma }q_{{\rm SW1-PY}}^{\prime }(0)=\left( b_{{\rm %
SHS1-PY}}-\frac{1}{24}\eta \lambda ^{2}\right) \sigma . 
\]%
On the other hand, if one takes the derivative of \ $q_{{\rm SHS1-PY}}(r)$
directly, then

\[
q_{{\rm SHS1-PY}}^{\prime }(0)=b_{{\rm SHS1-PY}}\ \sigma . 
\]

In conclusion, the difference in Eq. ($\ref{a4}$) stems from the following
one:

\begin{equation}
\lim_{R\rightarrow \sigma }q_{{\rm SW1-PY}}^{\prime }(0)=q_{{\rm SHS1-PY}%
}^{\prime }(0)-\frac{1}{24}\eta \lambda ^{2}\sigma .
\end{equation}

2) {\it MSA closure.}

The derivation of the MSA virial EOS is not plain and thus deserves some
remarks. In fact, one might believe that the SW-MSA virial EOS can be simply
obtained by inserting $g_{{\rm SW-MSA}}\left( R^{+}\right) -g_{{\rm SW-MSA}%
}(R^{-})=-\beta \epsilon $ into Eq. ($\ref{a2}$), which is exact for a
generic SW fluid. Unfortunately, this opinion is incorrect. In fact, as
\noindent pointed out by Smith {\it et al.} \cite{Smith77}, in the SW-MSA
case the procedure used for Eq. ($\ref{a2}$) becomes meaningless, because $%
y_{{\rm MSA}}(r)$ given by Eq. ($\ref{ymsa}$) is continuous at $r=\sigma ,$
but discontinuous at $r=R$ \ where the SW potential is discontinuous (on the
contrary, $y_{{\rm exact}}(r)$ is continuous everywhere). Consequently, Eq. (%
$\ref{a2}$) cannot be valid in the MSA.

Using the relationship

\[
\int_{A}^{B}dx\ F(x)\delta (x-x_{0})=\left\{ 
\begin{array}{cc}
\frac{1}{2}\left[ F\left( x_{0}^{-}\right) +F\left( x_{0}^{+}\right) \right]
, & \qquad \text{if \ \ \ }A<x_{0}<B, \\ 
0, & \text{otherwise,}%
\end{array}%
\right. 
\]%
which extends the usual definition of Dirac's $\delta $ to $F$-functions
discontinuous at $x_{0}$, one can demonstrate that the proper MSA expression
is{\it \ }\cite{Smith77}

\begin{eqnarray}
Z_{V}^{{\rm SW-MSA}} &=&1+4\eta \left\{ g_{{\rm SW-MSA}}(\sigma ^{+})-\beta
\varepsilon \ \left( R/\sigma \right) ^{3}\ \frac{1}{2}\left[ g_{{\rm SW-MSA}%
}(R^{-})+g_{{\rm SW-MSA}}(R^{+})\right] \right\}  \nonumber \\
&=&1+4\eta \left\{ 1+\gamma _{{\rm SW-MSA}}(\sigma )+\beta \varepsilon
-\beta \varepsilon \left( R/\sigma \right) ^{3}\left[ 1+\gamma _{{\rm SW-MSA}%
}(R)+\frac{1}{2}\beta \varepsilon \right] \right\} .
\end{eqnarray}%
For Baxter's model, the sticky limit of $Z_{V}^{{\rm SW1-MSA}}$ tends to $%
-\infty $, as $-\left( \beta \varepsilon _{{\rm SW1}}\right) ^{2}=\ln
^{2}\left( \alpha \ \frac{R}{R-\sigma }\right) .$

3) {\it C0-approximation.}

\ The C0-calculations are similar to the MSA ones. Now the resulting EOS is

\begin{eqnarray}
Z_{V}^{{\rm SW-C0}} &=&1+4\eta \left\{ 1+\gamma _{{\rm SW-C0}}(\sigma
)+e^{\beta \varepsilon }-1\right.  \nonumber \\
&&\left. -\beta \varepsilon \ \left( R/\sigma \right) ^{3}\left[ 1+\gamma _{%
{\rm SW-C0}}(R)+\frac{1}{2}e^{\beta \varepsilon }\right] \right\} .
\end{eqnarray}%
Unfortunately, also the sticky limit of $Z_{V}^{{\rm SW1-C0}}$ is infinite,
with the most divergent term being proportional to $-\beta \varepsilon _{%
{\rm SW1}}e^{\beta \varepsilon _{{\rm SW1}}}=-\alpha \ \frac{R}{R-\sigma }%
\ln \left( \alpha \ \frac{R}{R-\sigma }\right) $. 
%%%%%%%%%%%%%%%%%%%%%% Bibliography %%%%%%%%%%%%%%%%%%%%%%%%%%%%%%%%%%%%

%%%%%%%%%%%%%%%%%%%%% Figures %%%%%%%%%%%%%%%%%%%%%%%%%%%%%%%%%%%%%%%%%
\begin{figure}[tbp]
\caption{Structure factor $S(k \protect\sigma)$ versus dimensionless
wavector $k \protect\sigma$ for various closures ranging from MSA to PY, and
comparison with Monte Carlo results (MC) from Refs.~[43,44]. a) $\protect\tau%
=1.0$ and $\protect\eta=0.5$; b) $\protect\tau=0.5$ and $\protect\eta=0.4$ [
here, as well as in Fig. 2, the curves of the upper part are shifted upwards
by 2.5 units ].}
\label{Fig1}
\end{figure}
%\newpage 
\begin{figure}[tbp]
\caption{Same as in Fig.~1 with different temperatures and packing
fractions. c) $\protect\tau=0.2$ and $\protect\eta=0.32$; d) $\protect\tau%
=0.1$ and $\protect\eta=0.14$.}
\label{Fig2}
\end{figure}
%\newpage 
\begin{figure}[tbp]
\caption{Compressibility pressure $\protect\beta P_{C}\protect\sigma ^{3}$
as a function of the packing fraction $\protect\eta$ for two different
values of temperature (a) $\protect\tau=1.0$; b) $\protect\tau=0.5$) and
various closures [ here, as well as in all following figures for pressure
the curves of the upper part are shifted upwards by 2 units ].}
\label{Fig3}
\end{figure}
%\newpage 
\begin{figure}[tbp]
\caption{Same as in Fig.~3 with different temperatures c) $\protect\tau=0.2$%
; d) $\protect\tau=0.1$. }
\label{Fig4}
\end{figure}
%\newpage 
\begin{figure}[tbp]
\caption{Energy pressure $\protect\beta P_{E}\protect\sigma ^{3}$ as a
function of the packing fraction $\protect\eta$ for two different values of
temperature (a) $\protect\tau=1.0$; b) $\protect\tau=0.5$) and various
closures.}
\label{Fig5}
\end{figure}
%\newpage 
\begin{figure}[tbp]
\caption{Same as in Fig.~5 with different temperatures: c) $\protect\tau=0.2$%
; d) $\protect\tau=0.1$ .}
\label{Fig6}
\end{figure}
%\newpage 
\begin{figure}[tbp]
\caption{Structure factor $S(k \protect\sigma)$ versus dimensionless
wavector $k \protect\sigma$ for $\protect\tau=0.2$ and $\protect\eta=0.32$.
The performance of the GPY1 is tested against MC data of Ref.~[43] and PY
results.}
\label{Fig7}
\end{figure}
%%%%%%%%%%%%%%%%%%%%%%%%%%%%%%%%%%%%%%%%%%%%%%%%%%%%%%%%%%%%%%%%%%%%%%%%%%%%%%%


\begin{thebibliography}{910}

\bibitem{Stell91} G. Stell, J. Stat. Phys. {\bf 63}, 1203 (1991).

\bibitem{Jamnik96} A. Jamnik, J. Chem. Phys. {\bf 105}, 10511 (1996).

\bibitem{Santos98} A. Santos, S. B. Yuste, and M. L\'{o}pez de Haro, J.
Chem. Phys. {\bf 109}, 6814 (1998).

\bibitem{Jamnik01} A. Jamnik, J. Chem. Phys. {\bf 114}, 8619 (2001).

\bibitem{Baxter68} R. J. Baxter, J. Chem. Phys. {\bf 49}, 2270 (1968).

\bibitem{Baxter71a} R. J. Baxter, in: {\it Physical Chemistry, an Advanced
Treatise,} Vol. 8A, ed. D. Henderson (Academic Press, New York, 1971) ch. 4.

\bibitem{Baxter71b} R. O. Watts, D. Henderson and R. J. Baxter, Advan. Chem.
Phys. {\bf 21}, 421 (1971).

\bibitem{Barboy74} B. Barboy, J. Chem. Phys. {\bf 61}, 3194 (1974).

\bibitem{Perram75} J. W. Perram and E. R. Smith, Chem. Phys. Lett. {\bf 35},
138 (1975).

\bibitem{Barboy79} B. Barboy and R. Tenne, Chem. Phys. {\bf 38}, 369 (1979).

\bibitem{Brey87} J. J. Brey, A. Santos, and F. Casta\~{n}o, Mol. Phys. {\bf %
60}, 113 (1987).

\bibitem{Mier89} L. Mier-y-Teran, E. Corvera, and A. E. Gonzalez, Phys. Rev.
A {\bf 39}, 371 (1989).

\bibitem{Ginoza96} M. Ginoza and M. Yasutomi, Mol. Phys. {\bf 87}, 593
(1996).

\bibitem{Tutschka98} C. Tutschka and G. Kahl, J. Chem. Phys. {\bf 108}, 9498
(1998).

\bibitem{Tutschka00} C. Tutschka and G. Kahl, Phys. Rev. E {\bf 62}, 3640
(2000).

\bibitem{Tutschka01} C. Tutschka and G. Kahl, Phys. Rev. E {\bf 64}, 031104
(2001).

\bibitem{Tutschka02} C. Tutschka, G. Kahl, and E. Riegler, Mol. Phys. {\bf %
100}, 1025 (2002).

\bibitem{Gazzillo00} D. Gazzillo and A. Giacometti, J. Chem. Phys. {\bf 113}%
, 9846 (2000).

\bibitem{Gazzillo02a} D. Gazzillo and A. Giacometti, Physica A {\bf 304},
202 (2002).

\bibitem{Gazzillo02b} D. Gazzillo and A. Giacometti, Mol. Phys. {\bf 100},
3307 (2002).

\bibitem{Gazzillo03a} D. Gazzillo and A. Giacometti, J. Appl. Cryst.{\it \ }%
{\bf 36}, 832 (2003).

\bibitem{Gazzillo03b} D. Gazzillo and A. Giacometti, Mol. Phys. {\bf 101},
2171 (2003).

\bibitem{Timoneda89} J. Juan\'{o}s i Timoneda and A. D. J. Haymet, Phys.
Rev. A {\bf 40}, 5979 (1989).

\bibitem{Ginoza01} M. Ginoza, Mol. Phys. {\bf 99}, 1613 (2001).

\bibitem{Jamnik91} A. Jamnik, D. Bratko and D. J. Henderson, J. Chem. Phys. 
{\bf 94}, 8210 (1991).

\bibitem{Hoye93} J. S. H\o ye, E. Lomba, and G. Stell, Mol. Phys. {\bf 79},
523 (1993).

\bibitem{Huang84} J. S. Huang, S. A. Safran, M. W. Kim, G. S. Grest, M.
Kotlarchyk, and N. Quirke, Phys. Rev. Lett. {\bf 53}, 592 (1984).

\bibitem{Pini02} D. Pini, A. Parola, and L. Reatto, Mol. Phys. {\bf 100},
1507 (2002).

\bibitem{Rasaiah85} J. C. Rasaiah and S. H. Lee, J. Chem. Phys. {\bf 83},
6396 (1985).

\bibitem{Herrera91} J. N. Herrera and L. Blum, J. Chem. Phys. {\bf 94}, 5077
(1991).

\bibitem{note0} We adopt the following terminological distinction. A SHS 
{\it model} is univocally defined by its Hamiltonian, whose specification
requires {\it two} elements: the starting potential, and the `sticky limit'
procedure. On the other hand, a particular {\it solution }is characterized
by {\it three} elements: the same two of the model, plus an approximate
`closure' for the OZ equation.

\bibitem{Katsov00} K. Katsov and J. D. Weeks, J. Stat. Phys. {\bf 100}, 107
(2000).

\bibitem{Stell63} G. Stell, Physica {\bf 29}, 517 (1963).

\bibitem{Post86} A. J. Post and E. Glandt, J. Chem. Phys. {\bf 84}, 4585
(1986).

\bibitem{Smith77} W. R. Smith, D. Henderson and Y. Tago, J. Chem. Phys. {\bf %
67}, 5308 (1977).

\bibitem{note1} Here, $\ \delta _{+}(x)$ is the asymmetrical Dirac delta
function defined by: $\int_{A}^{B}dx\ F(x)\delta _{+}(x-x_{0})=F\left(
x_{0}^{+}\right) ,$ if $A\leq x_{0}<B$, and \ $=0,$ if $x_{0}<A$ or $%
x_{0}\geq B.$

\bibitem{Henderson75} D. Henderson and E. W. Grundke, J. Chem. Phys. {\bf 63}%
, 601 (1975).

\bibitem{Regnaut89} C. Regnaut and J. C. Ravey, J. Chem. Phys. {\bf 91},
1211 (1989).

\bibitem{Seaton87} N. A. Seaton and E. D. Glandt, J. Chem. Phys. {\bf 86},
4668 (1987).

\bibitem{Yuste93a} S. Bravo Yuste and A. Santos, J. Stat. Phys. {\bf 72},
703 (1993).

\bibitem{Yuste93b} S. Bravo Yuste and A. Santos, Phys. Rev. E {\bf 48}, 4599
(1993).

\bibitem{Kranendonk88} W. G. T. Kranendonk and D. Frenkel, Mol. Phys. {\bf 64%
}, 403 (1988).

\bibitem{Miller03} M. A. Miller and D. Frenkel, private communication.

\bibitem{Frenkel03} M. A. Miller and D. Frenkel, Phys. Rev. Lett. {\bf 90},
135702\ \ (2003)

\end{thebibliography}
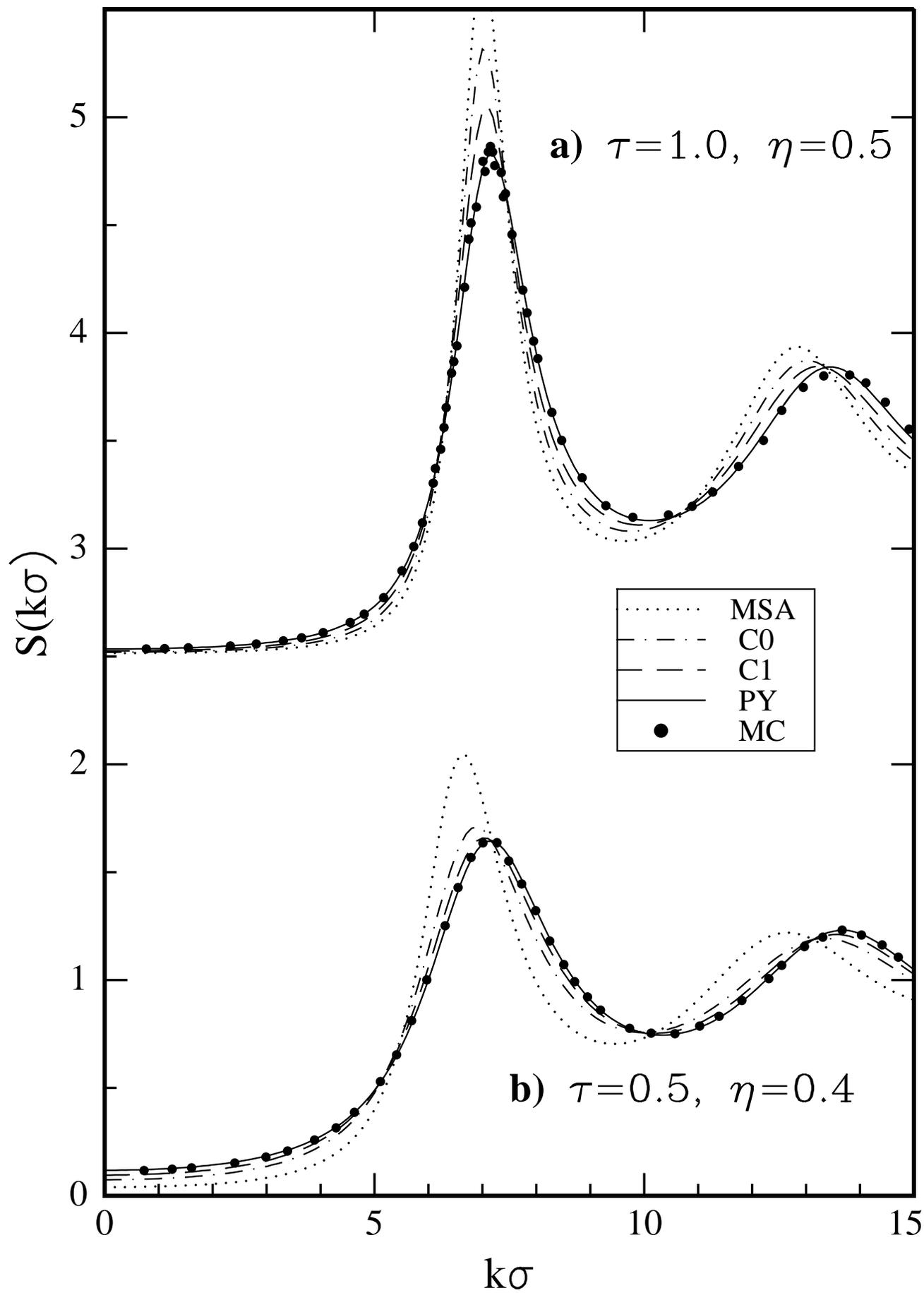
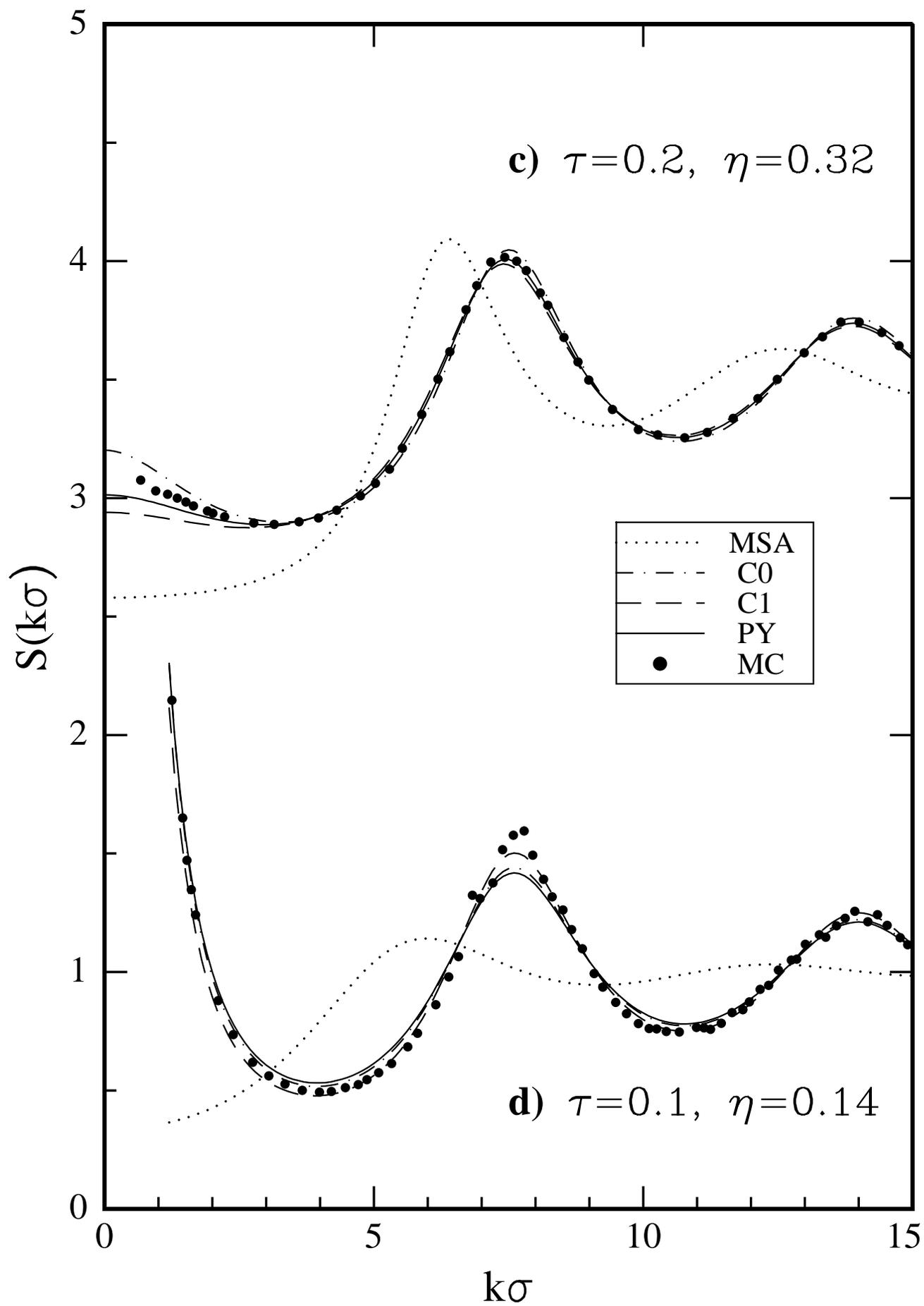
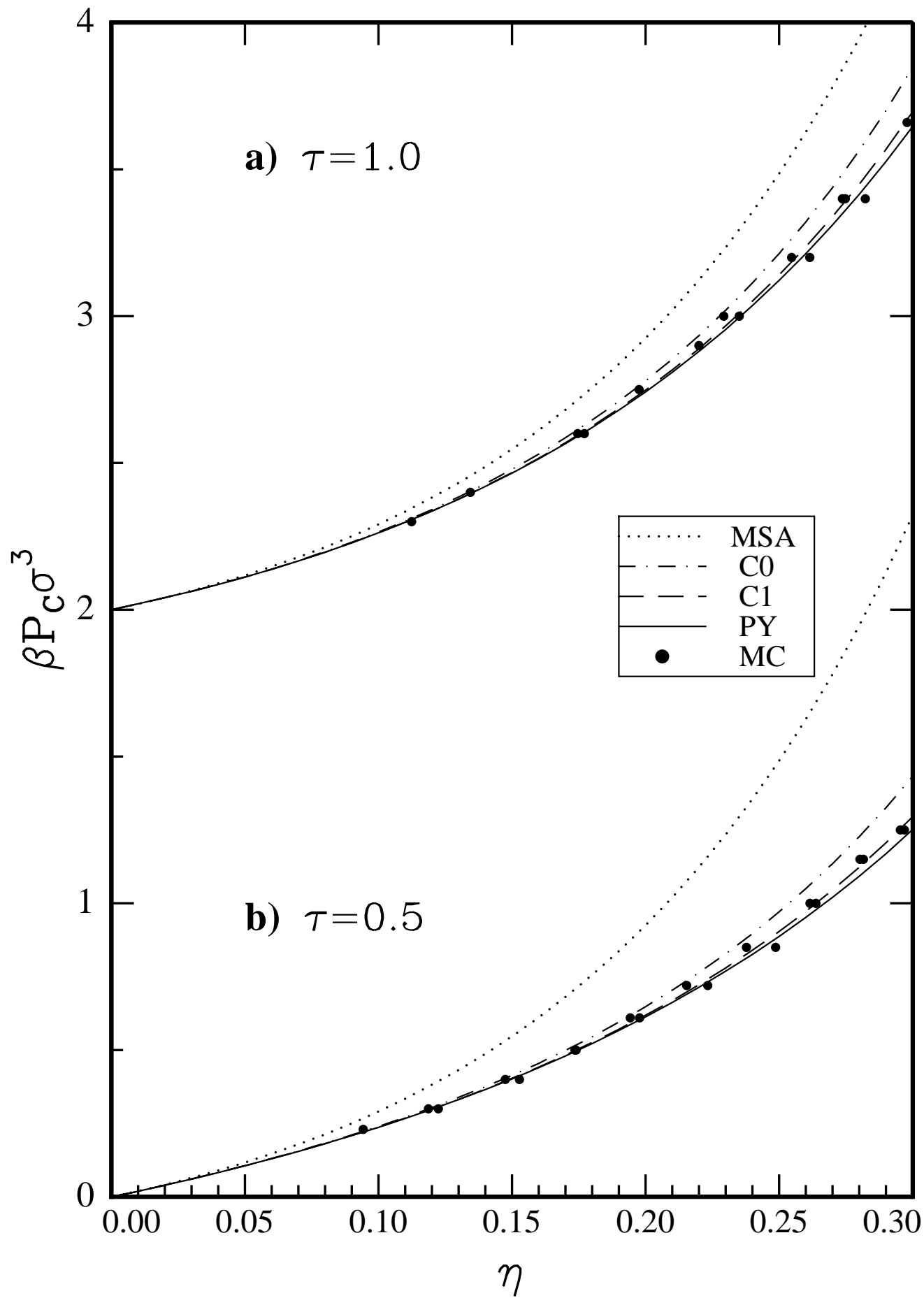
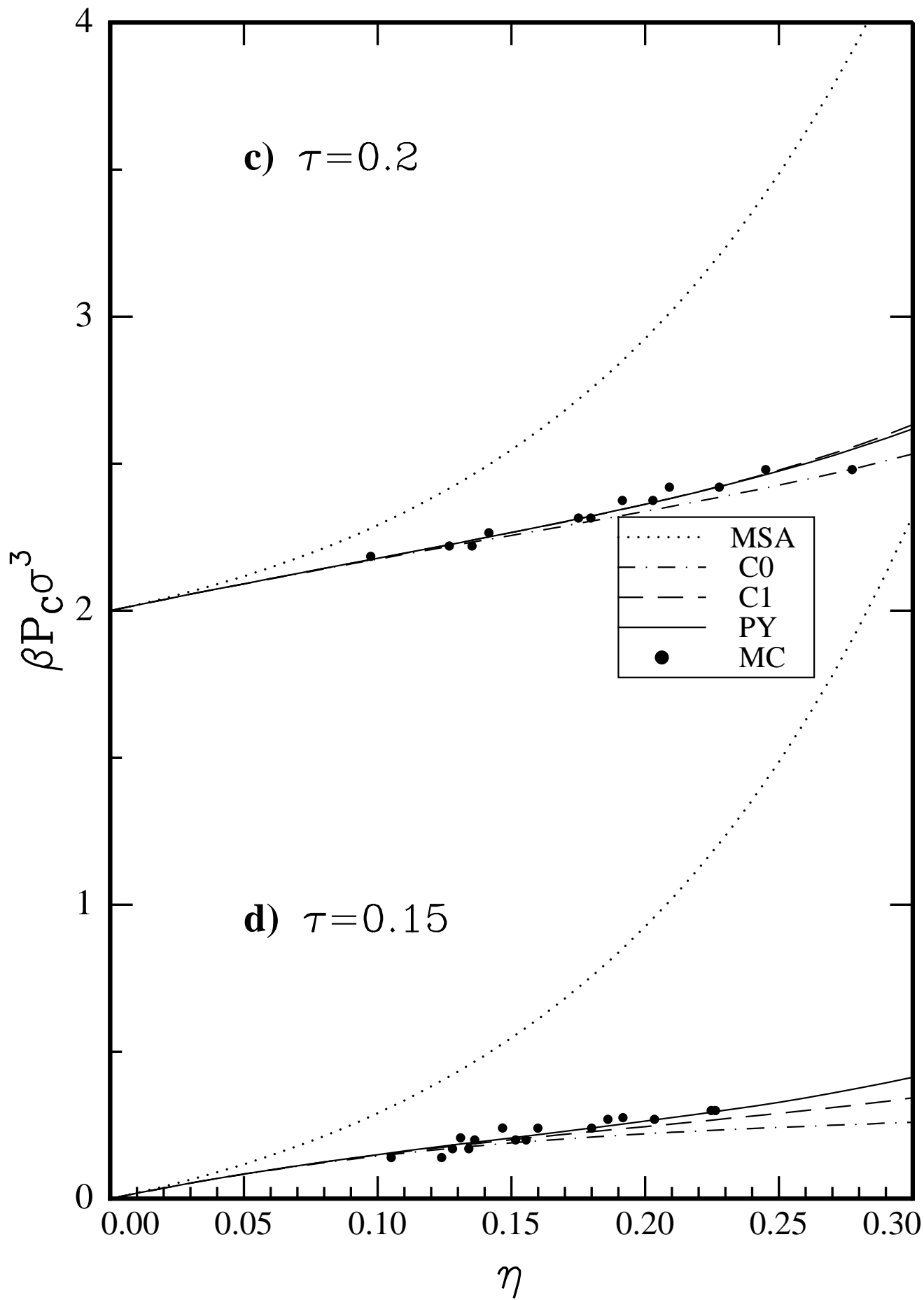
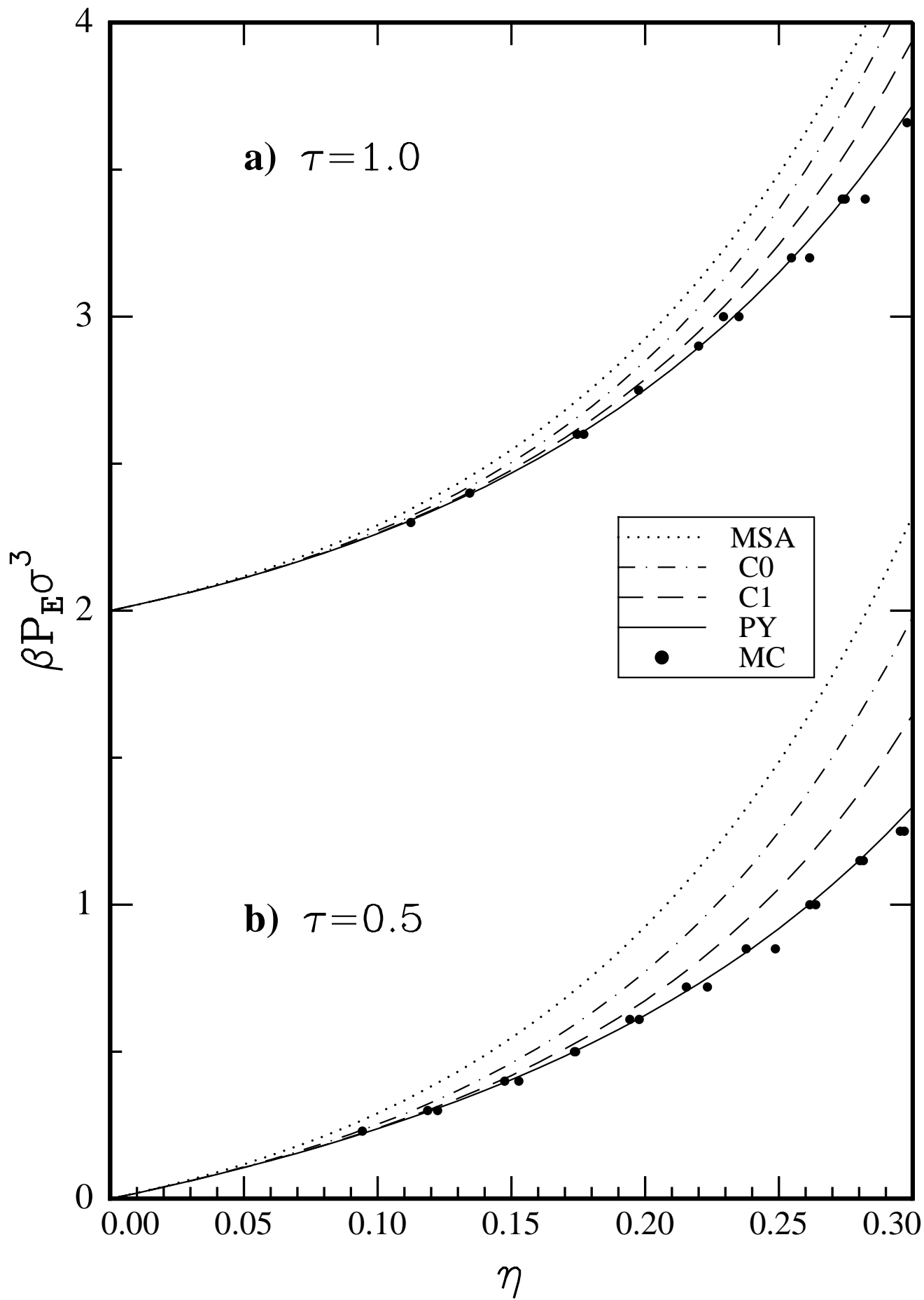
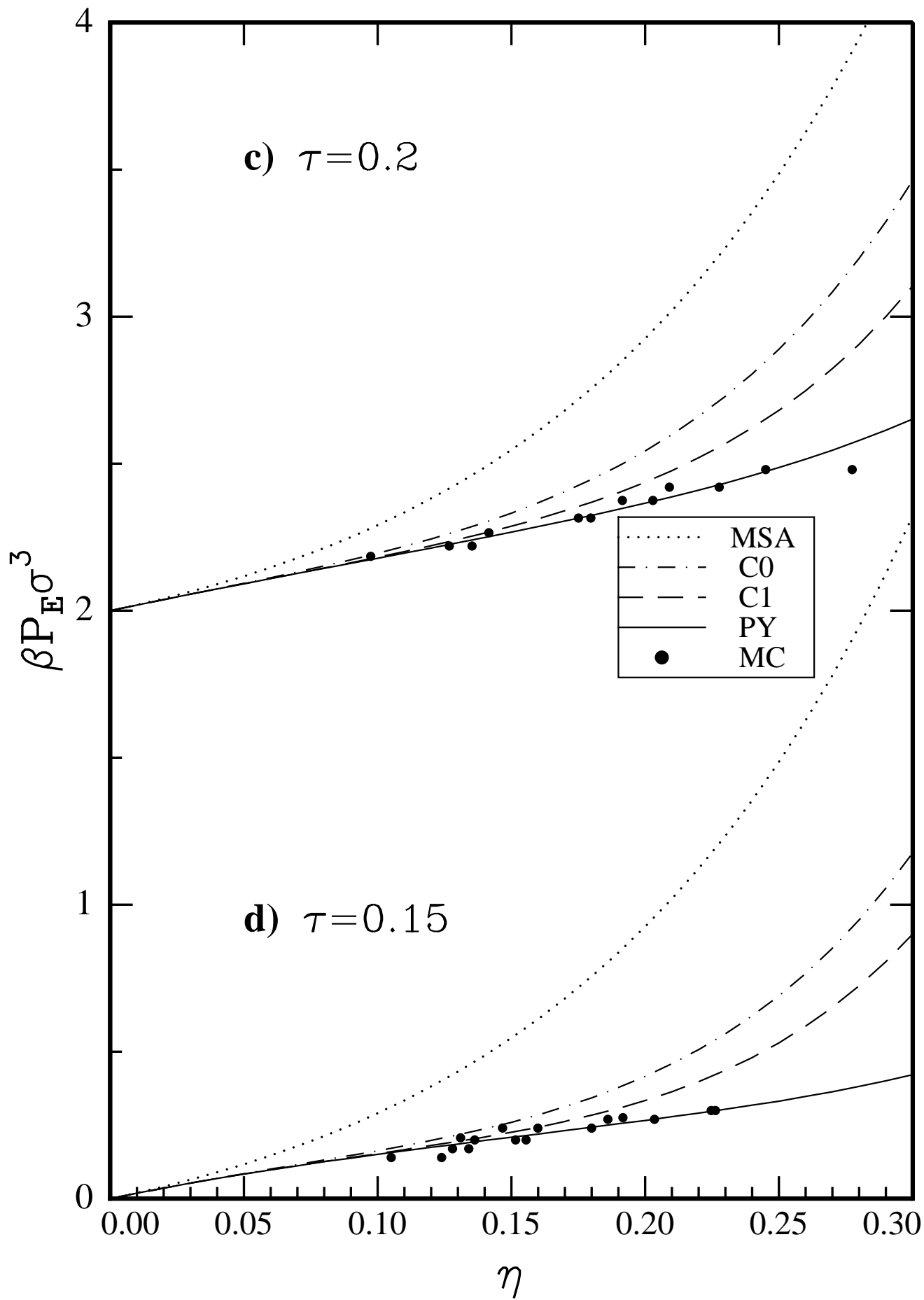
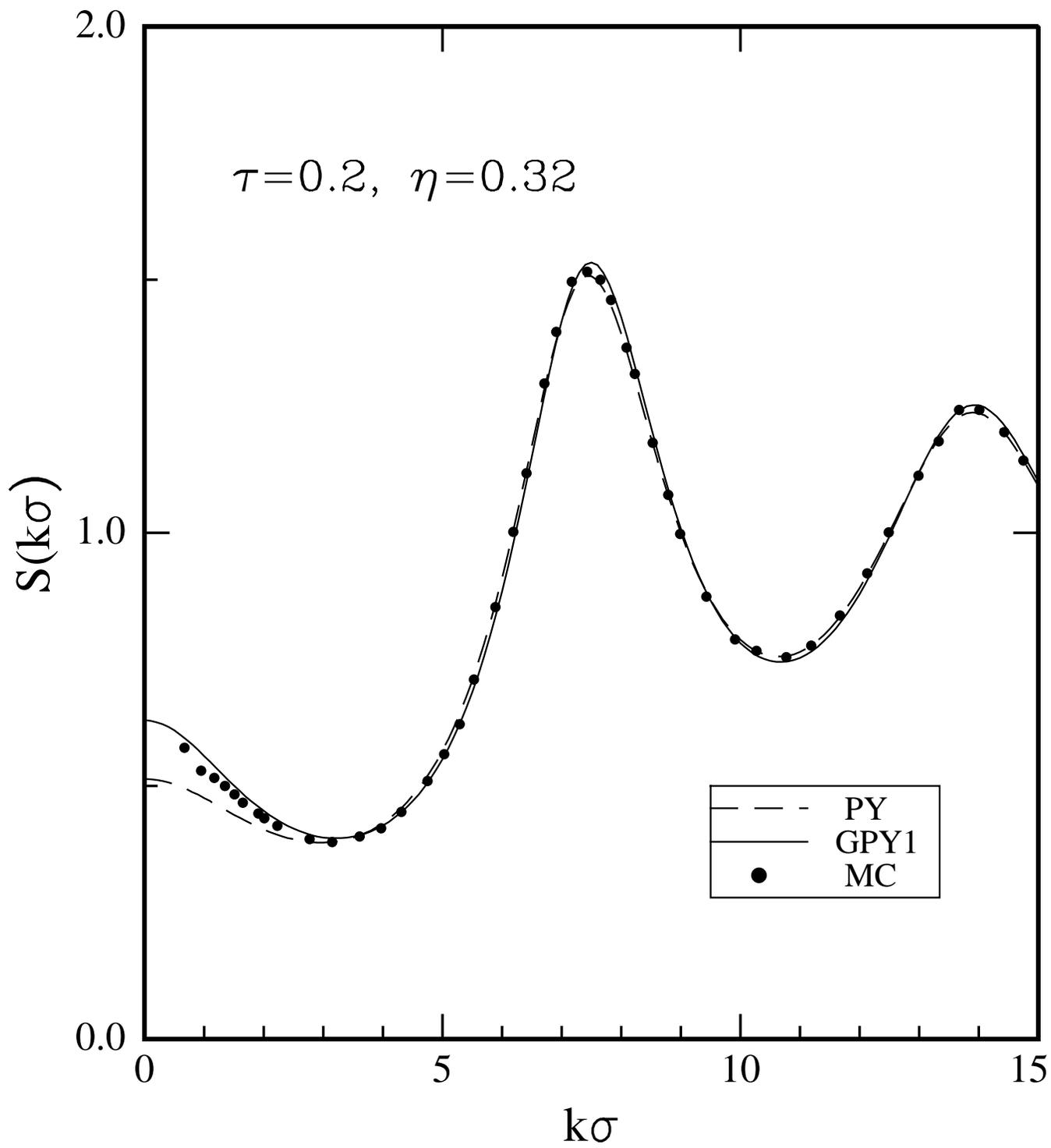
\end{document}